\documentclass[useAMS,usenatbib]{mn2e}
\usepackage{times}
\usepackage{graphicx}
\usepackage{amssymb}
\usepackage[usenames,dvipsnames]{color}
\newcommand\herschel{\textit{Herschel}}
\newcommand\microns{$\mu$m}

\title[Overdensity near MRC\,1138-26]{Serendipitous detection of an overdensity of \textit{Herschel}-SPIRE 250 \micron\ sources south of MRC\,1138-26\thanks{\herschel\ is an ESA space observatory with science instruments provided by European-led Principal Investigator consortia and with important participation from NASA.}}
\author[I. Valtchanov et al.]
{
Ivan Valtchanov$^{1}$\thanks{E-mail:ivaltchanov@sciops.esa.int},
B. Altieri$^{1}$,
S. Berta$^{2}$,
E. Chapin$^{3}$,
D. Coia$^{1}$, 
L. Conversi$^{1}$,\newauthor 
H. Dannerbauer$^{4}$,
H. Dom\'inguez-S\'anchez$^{5,6}$,
T. D. Rawle$^{1}$,
M. S\'anchez-Portal$^{1}$,\newauthor
J.\,S. Santos$^{1}$,
S. Temporin$^{7}$,\\
$^{1}$Herschel Science Centre, European Space Astronomy Centre, ESA, 28691 Villanueva de la Ca\~nada, Spain\\
$^{2}$Max-Planck-Institut f\"{u}r Extraterrestrische Physik (MPE), Postfach 1312, 85741, Garching, Germany\\
$^{3}$XMM-Newton SOC, European Space Astronomy Centre, ESA, 28691 Villanueva de la Ca\~nada, Spain\\
$^{4}$Universit\"at Wien, Institut f\"ur Astrophysik, T\"urkenschanzstra{\ss}e 17, 1180 Wien, Austria\\
$^{5}$Instituto de Astrof\'{\i}sica de Canarias, 38200 La Laguna,
Tenerife, Spain\\
$^{6}$Departamento de Astrof\'{\i}sica, Universidad de La Laguna,38206
La Laguna, Tenerife, Spain\\
$^7$Institute for Astro- and Particle Physics, University of Innsbruck,
Technikerstra{\ss}e 25, A-6020, Austria
}

\begin{document}

\date{Accepted 2013 September 14. Received 2013 September 13; in original form 2013 April 17}

\pagerange{\pageref{firstpage}--\pageref{lastpage}} \pubyear{2013}

\maketitle

\label{firstpage}

\begin{abstract}
We report the serendipitous detection of a significant overdensity of \herschel-SPIRE 250 \microns\ sources in the vicinity of MRC\,1138-26. We use an adaptive kernel density estimate (AKDE) to quantify the significance, including a comparison with other fields. The overdensity has a size of $\sim3\farcm5$-$4\arcmin$ and stands out at $\sim5\sigma$ with respect to the background estimate. No features with similar significance were found in four \textit{Herschel}-observed extragalactic control fields: GOODS-North, Lockman, COSMOS and UDS. The chance of having a similar overdensity in {a field with the same number but randomly distributed sources} is less than 2\%. The clump is also visible as a low surface brightness feature in the \textit{Planck} 857 GHz map. We detect 76 sources at 250 \micron\ (with a signal-to-noise ratio greater than 3), in a region of $4\arcmin$ radius (or $1.51\pm0.17$ arcmin$^{-2}$); 43 of those (or $0.86\pm0.13$ arcmin$^{-2}$) are above a flux density limit of 20 mJy.  This is a factor of 3.6 in excess over the average $0.24\pm0.02$ arcmin$^{-2}$ in the four control fields, considering only the sources above 20 mJy.  We also find an excess in the number counts of sources with 250 \micron\ flux densities between 30 and 40 mJy, compared to deep extragalactic blank-field number counts. Assuming a fixed dust temperature (30 K) and emissivity ($\beta=1.5$) a crude, blackbody-derived redshift distribution, $z_\mathrm{BB}$, of the detected sources is significantly different from the distributions in the control fields and exhibits a significant peak at $z_\mathrm{BB} \approx 1.5$, although the actual peak redshift is highly degenerate with the temperature. We tentatively suggest, based on $z_\mathrm{BB}$ and the similar S250/S350 colours of the sources within the peak, that a significant fraction of the sources in the clump may be at a similar redshift. {Since the overdensity lies $\sim7\arcmin$ south of the $z=2.16$ Spiderweb protocluster MRC\,1138-26, an intriguing possibility (that is presently unverifiable given the data in hand) is that it lies within the same large-scale structure.}
\end{abstract}

\begin{keywords}
galaxies: clusters: general --- galaxies: high-redshift --- galaxies: individual: MRC\,1138$-$26 --- infrared: galaxies
\end{keywords}

\section{Introduction}

Star-formation in the local universe occurs mainly in late-type galaxies and they are predominantly found in the field, in loose galaxy groups, or in the infall regions of rich clusters of galaxies (e.g. \citealt{haines09}). At higher redshift the situation is more complicated and there are reports of the so-called ``reversal'' of the star-formation density relation (e.g. \citealt{popesso11}). At epochs near the peak of the star-formation rate density history (at redshifts 2--3; e.g. \citealt{hopkins06}), the massive clusters of today are not yet formed or virialised, and the so called protoclusters are still accreting material from the surroundings, which in most cases include dusty star-forming galaxies (DSFG). As predicted by \citet{granato04}, following the standard hierarchical model of galaxy formation, the spheroids falling into the massive halo will have a peak of star-formation occurring at more or less the same epoch. Consequently, we expect to find overdensities of DSFGs, especially in protocluster regions. {Indeed, a number density excesses of DSFGs are found around distant radio sources and QSOs, with overdensity sources identified using Ly$\alpha$ and other rest-frame UV or optical emission or absorption lines (e.g. \citealt{kurk04a,kurk04b,venemans07,kuiper11}), optical/NIR colour selections (e.g. \citealt{kajisawa06,kodama07,hatch11}) or submm imaging (e.g. \citealt{stevens10}). As the distant QSOs and the powerful radio galaxies are considered to be the likely progenitors of the brightest cluster galaxies in the local Universe (e.g. \citealt{ivison13} and the references therein) consequently these studies are important for our understanding of the build-up of today's massive clusters of galaxies. However, none of the reported excesses can be linked to any visually remarkable overdensity of sources in broad band maps. }

In addition to the DSFG excesses found in many protocluster fields, it is also well established that the submm detected star-forming galaxies (SMGs;  $\sim1$ mm-detected subset of DSFGs) trace the densest large-scale structures in the high-redshift universe  (e.g. \citealt{chapman09}) and are strongly clustered, with a characteristic correlation length of the order of 5--15 Mpc (e.g. \citealt{blain04,scott06, weiss09,hickox12}). These studies are usually limited by a small area, and hence low number statistics, which were, {however,} overcome with the \herschel\ Space Observatory \citep{pilbratt10} and the Spectral and Photometric Imaging REceiver (SPIRE, \citealt{griffin10}). 

SPIRE can quickly map large areas of the sky in three bands simultaneously (250, 350 and 500 \micron) down to confusion limited depths of (5.8,6.3,6.8) mJy \citep{nguyen10} and with a diffraction-limited resolution of (18,24,35)$\arcsec$  beam FWHM in the three bands, respectively \citep{griffin13}. The spectral energy distributions (SEDs) of star-forming galaxies peak at $\sim100$ \micron\ (rest-frame), which means that the majority of SPIRE sources are high-redshift $z > 1$ star-forming galaxies.  Down to the confusion limit at 250 \micron, there are on average $\sim 3700$ galaxies deg$^{-2}$ (see \citealt{hermes}, Table 7). There are a number of SPIRE surveys exploring the high redshift Universe, but only two cover fields large enough to allow studies of DSFG clustering, the \textit{Herschel} Multi-tiered Extragalactic Survey (HerMES, \citealt{hermes}) and the \textit{Herschel} Astrophysical Terahertz Large Area Survey (H-ATLAS, \citealt{hatlas}). Both report similar results on the correlation length, compatible with the earlier SMG studies: 7--11 Mpc for the shallower H-ATLAS \citep{maddox10}, and 5--6 Mpc for HerMES \citep{cooray10}. These studies use resolved sources in order to estimate the clustering; however they do not actually identify significant overdensities.

An efficient way to detect clumps of strongly clustered sources, like the DSFGs, is to observe with an instrument with a spatial resolution (the beam) that matches their correlation length. This was developed in \citet{negrello05} with a particular emphasis on \textit{Planck} and \textit{Herschel}. \textit{Planck} beam sizes range from $5\arcmin$ at 857 GHz  to $30\arcmin$ at 100 GHz \citep{planck} {and they better match} the DSFG correlation length of 7 Mpc (or $14\arcmin$ at $z=2$) than \textit{Herschel}-SPIRE. Indeed, follow-up of 16 \textit{Planck} early release compact sources revealed that in 4 cases these resolved into overdensities of faint SPIRE galaxies \citep{clements13}. There are more similar studies to come (\textit{Planck} collaboration, in preparation) exploring the all-sky coverage of \textit{Planck}, although due to the particular frequency coverage, the samples will be biased to clumps of dusty galaxies at $z>2$. 

In this paper we report the detection of a significant and visually compelling overdensity of SPIRE 250 \micron\ detected galaxies. This is a pilot study of overdensities found in extragalactic \textit{Herschel} maps. Note that throughout this paper, when we refer to overdensities, we mean a visual overdensity of sources in maps. We intentionally do not refer to these clumps as ``candidate protoclusters'' of dusty galaxies as they could just be chance alignments.

The structure of the paper is the following: in the next section (\S~\ref{sec:data}) we introduce the \textit{Herschel}-SPIRE observations and the data processing, giving estimates of the map depth and sensitivity, the source detection procedure, the source catalogue statistics and completeness and we introduce the four extragalactic control fields used in the analysis. Then in \S~\ref{sec:over} we proceed by presenting the adaptive kernel density estimate method employed to characterise the overdensity and the different strategies used to estimate its significance. In \S~\ref{sec:surf} we present the surface density measurements in comparison with the control fields and in \S~\ref{sec:flux} we do the same for the distribution of the 250 \micron\ source flux densities. Next, in \S~\ref{sec:zbb}, we estimate their black-body derived redshifts and {their distribution in a colour-magnitude diagram}. The {corresponding detection of the overdensity} by \textit{Planck}-HFI at 857 GHz is briefly discussed in \S~\ref{sec:planck}. The conclusions are presented in \S~\ref{sec:conclusions}.

We use the following cosmological parameters: $\Omega_m = 0.3$, $\Omega_{\lambda} = 0.7$, $H_0 = 70$ km s$^{-1}$ Mpc$^{-1}$.

\section{Observations and data processing}
\label{sec:data}

The \herschel-SPIRE and Photodetector Array Camera and Spectrometer (PACS, \citealt{pacs}) observations were performed as part of \herschel\ guaranteed time (PI: B. Altieri). The target of the \herschel\ observations is a well known protocluster: MRC\,1138-26, also known as the Spiderweb protocluster. Its {existence} was confirmed by reports of significant overdensities of Ly\,$\alpha$, H\,$\alpha$ emitters and near-IR galaxy colours consistent with 
the 4000 \AA\ break at $z\sim2.2$, close to the redshift $z=2.16$ of the Spiderweb radio galaxy (e.g. \citealt{pentericci00,kurk04a,kurk04b,miley06,venemans07}). The SPIRE and PACS observations were used in a detailed analysis of the central massive star-forming radio galaxy \citep{seymour12}. 

Comprehensive details of the \textit{Herschel} data processing, map-making and source detection, and catalogues of the \herschel\ observations, will be presented in S\'anchez-Portal et al. (in preparation).

\subsection{SPIRE map making}
The SPIRE observation (\textsc{obsid}: 1342210877, on 07 Dec 2010) was executed in Large Map mode covering a field of $\sim20\arcmin \times 20\arcmin$, centred on the radio galaxy (RA=11:h40:48.360, Dec=-26:29:10.60, J2000.0). To increase the depth, the same area was observed four times in sequence (i.e. four map repetitions). Maps were produced with the \herschel\ Interactive Processing Environment (HIPE v10\footnote{HIPE \citep{hipe} is a joint development by the \herschel\ Science Ground Segment Consortium, consisting of ESA, the NASA \herschel\ Science Center, and the HIFI, PACS and SPIRE consortia}), following the standard data processing and map-making steps, using \textsc{spire\_cal\_10\_1} calibration tree. The final maps have a resolution of  (6, 10, 14)$\arcsec$ per pixel at (250, 350, 500) \micron. 

\begin{figure}
\centerline{
\includegraphics[width=8cm]{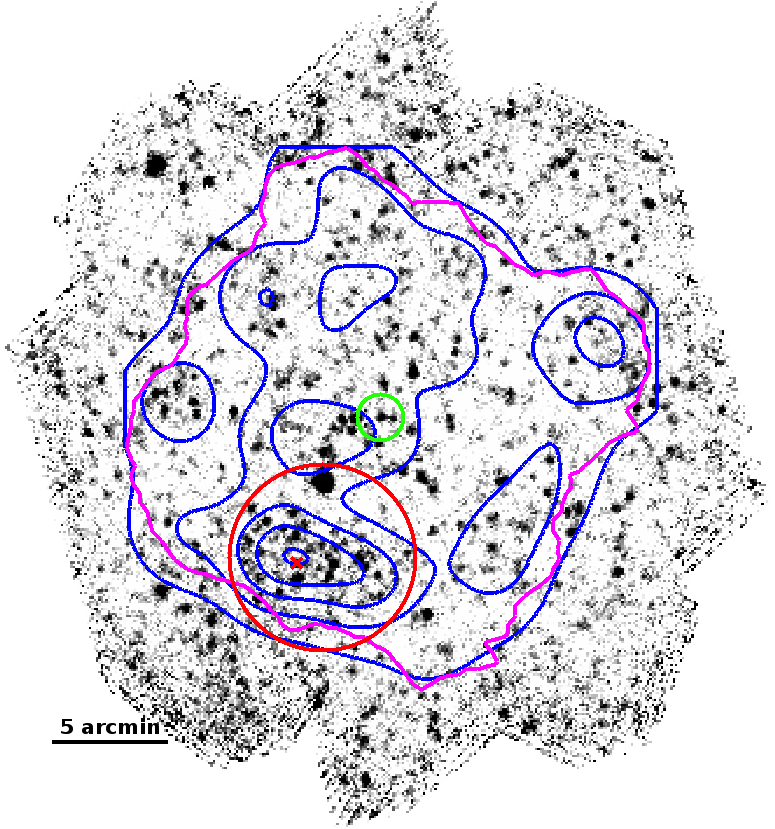}}
\caption{SPIRE 250 \micron\ image with blue contours showing the AKDE of the galaxies detected at 250 \micron. The contour levels are from 1 to 5$\,\sigma$ with respect to the background AKDE \textit{rms} noise. The source with the peak AKDE of $6\sigma$ is marked with a red ``$\times$'' sign. The Spiderweb galaxy MRC\,1138-26 is indicated as a $1\arcmin$ green circle. The distance between MRC\,1138-26 and the peak AKDE is $\sim7\arcmin$, which corresponds to a comoving distance of $\sim10$ Mpc, if the overdensity is part of the Spiderweb protocluster at $z\sim 2.2$. The magenta contour delineates the region with relative coverage above 50\%. For the AKDE we use detections only inside this area of $\sim 17\arcmin \times 17\arcmin$. The red circle, centred on RA=11:40:59.5, Dec=-26:35:11, of $4\arcmin$ radius, shows the overdensity region that we consider for the analysis (see Section~\ref{sec_over_def}). North is up and East to the left. }
\label{fig_over}
\end{figure}

Figure~\ref{fig_over} shows the SPIRE 250 \micron\ map with the \textit{Herschel} telescope turnaround data included. We include the turnaround data for the map making as it increases the area, but the coverage in these turnaround regions is quite irregular, and this affects the source detection homogeneity. We have therefore not used it in the subsequent analysis. We corrected the astrometry of the final maps using \textit{Spitzer\/}-MIPS 24 \micron\ and \herschel-PACS 100 and 160 \micron\ maps, which are only available for the central few arcmin of the \mbox{MRC\,1138-26} field. The overall astrometry offset was less than $2\arcsec$, consistent with \herschel's absolute pointing error \citep{pilbratt10}. Our estimate of the residual astrometry offset, after the correction, is at the sub-arcsec level. 

\subsection{Map depth and sensitivity}
\label{rms}
The median number of bolometer readouts in the central area of the map, where the relative coverage is higher than 50\%, is (68,115,124) at (250,350,500) \micron\, which corresponds to an effective exposure time of (4,6,7) seconds with the SPIRE photometer bolometer readout frequency of 18.3 Hz. The \textit{rms} noise of the final maps in the same central region is (6.8, 8.3, 7.6) mJy at (250, 350, 500) \microns\ correspondingly, derived using the robust median absolute deviation (MAD), multiplied by 1.4826 to match the \textit{rms} for a nominally Gaussian noise. This is a good approximation for SPIRE, as the  maps have little 1/f noise \citep{nguyen10}. These \textit{rms} levels are slightly above the SPIRE extragalactic confusion limits, because some contribution from sources may still affect the MAD measurement. 

\subsection{Source detection and completeness}

We use the \textsc{Sussextractor} (SXT) source detection method developed for SPIRE maps (see \citealt{smith12} for more details). The SXT is a maximum likelihood method that treats each potential source (e.g. the centre of each pixel) and considers two models at each location: an empty sky model and a source model. The two models are compared in order to determine whether a source is present at the particular location. The method assumes the source model is known, i.e. the beam profile for SPIRE. There are 817 sources detected at a signal-to-noise ratio (SNR) above 3 in the SPIRE 250 \micron\ map of the Spiderweb field, 366 of those are within the region of relative coverage greater than 50\% (indicated with a magenta contour in Figure~\ref{fig_over}). We use the SPIRE 250 \micron\ detected sources as an input prior list for the maximum likelihood fits at 350 and 500 \micron.  The prior entries that result in a poor fit at 350 or 500 \micron\ are replaced with $3\sigma$ upper limits based on the nominal confusion noise (i.e. 19 and 20 mJy respectively, see Sect.~\ref{rms}). Note that a single source at 250 \micron\ may be multiple galaxies, but without shorter wavelength data we cannot flag this blending (see e.g. \citealt{elbaz11}).  By using 250 \micron\ detections as priors we may miss very high redshift sources ($z>3$). As our maps are confusion noise dominated we can {refer to} the completeness estimates from other confusion dominated maps, for example GOODS-North from HerMES \citep{smith12}. Hence the flux density at 50\% completeness levels are approximately (12,13,13) mJy at (250,350,500) \micron. 

\subsection{Extragalactic control fields}
\label{sec_fields}

{We have chosen four well studied extragalactic fields to use as control fields: GOODS-North, COSMOS, Lockman, and UDS.} These fields are part of HerMES (see \citealt{hermes} for more details on field locations and the \herschel\ observations). The SPIRE maps in these fields, shown in Figure~\ref{fig_fields}, are larger ($30\arcmin\times30\arcmin$) and much deeper than the Spiderweb observation, so we have reprocessed them by limiting to the same depth as the Spiderweb map, i.e. only using four map repetitions. Subsequently we apply the same map making and source detection procedure. In order to make sure that we match the depth of the Spiderweb observation we compare the derived $1\sigma$ noise in the 250 \micron\ maps of the control fields: using the robust MAD estimator in a similar way as in Sect.~\ref{rms} we obtain slightly lower but consistent value of 6 mJy in GOODS-North, COSMOS, UDS and 5.6 mJy in Lockman.

\section{An overdensity of SPIRE sources}
\label{sec:over}

\subsection{Adaptive kernel density estimate (AKDE)}

A visual overdensity of sources south-east of the Spiderweb galaxy stands out clearly with respect to the field in the 250 \micron\ map (Figure~\ref{fig_over}). The feature has a low contrast at 350 and 500 \micron. In order to quantitatively define the overdensity and assess its significance, we apply an iterative adaptive kernel density estimate on the detected 250 \micron\ galaxies. We follow \citet{pisani96} to compute the optimal Gaussian kernel width at any point within the map, based on the local density of detected sources. The method of \citet{pisani96} is iterative and the optimal kernel width at each point is obtained after minimisation of a suitable function (the integrated square error of the true density and the estimated one, see \citealt{pisani96} for more details) using only the positions of the sources. \citealt{ferdosi11} provides more details and a comprehensive comparison of the different methods for density estimation.

The AKDE for the Spiderweb field is derived using all 366 detections at 250 \micron\ within the area with coverage above 50\%. The procedure provides the density estimate and the optimal Gaussian width at each input source. The maximum observed signal-to-noise density is $6.0\sigma$, on a source indicated by a cross sign in Figure~\ref{fig_over}, where $\sigma$ is taken as the standard deviation (st.dev.) of all 366 density estimates. The density and the optimal Gaussian width at each observed source are then used to make an AKDE map on a 301 by 301 pixels grid (similar to the observed 250 \micron\ map of $328\times356$ including telescope turnaround data). As expected, the peak density on the map is smaller -- at $5.1\sigma$ -- as the gridding smooths out the signal. The AKDE result is shown as contours from 1 to 5$\sigma$ in Figure~\ref{fig_over}.

\subsection{AKDE peak significance relative to random fields} 

To properly quantify the significance of the overdensity we generate the same number of galaxies with random positions within the 50\% map coverage area (the magenta contour in Figure~\ref{fig_over}) and we impose a minimum nearest neighbour distance of 2 pixels (or 12\arcsec\ for the 250 \micron\ map). This minimum distance is chosen to mimic the actual empirical observing selection by which distinct objects within 2/3 of the $18\arcsec$ beam FWHM at 250 \micron\  cannot be separated. 

For each one of the 100 random realizations we derive the AKDE on a $301\times301$ pixels maps.  In 2 out of 100 random realisations we observe a peak AKDE that exceeds the observed one of $5.1\sigma$. The maximum random field peak is at $5.4\sigma$ with AKDE {maximum} values ranging from $3.5\sigma$ to $5.4\sigma$. This means that the probability of finding a similar overdensity in a random field is less than 2\%.  Note that 2\% is an upper limit, because in our estimate we only take into account the peak value of the AKDE map and we do not consider the actual size of the region { -- for example, a small number of closely packed sources would produce a greater AKDE value, but the AKDE contour above 3 or 4$\sigma$ (as a measure of the size) would be much smaller in this case.}

\subsection{AKDE peak significance relative to control fields} 

Using the SXT catalogues in the four control fields (see Section~\ref{sec_fields}) we calculated the AKDE following the same procedure and the results are shown in Figure~\ref{fig_fields}.  In GOODS-N, COSMOS and UDS the highest AKDE peaks are around 3.2--3.6$\,\sigma$. The most significant peak, at $4.9\,\sigma$, is found in the Lockman field, with a number of large ($\gtrsim 10\arcmin$) irregular regions at 4$\,\sigma$. These Lockman features are interesting, but we defer their analysis to a future work. As indicated in Figure~\ref{fig_fields} none of the $3\sigma$ or above regions can be linked to known protoclusters \citep{papovich10,chapman09} or cluster candidates at $z>1$ \citep{wen11}.

\subsection{Definition of the overdensity}
\label{sec_over_def}

The AKDE peak at $5.1\sigma$ in the field of MRC\,1138-26, is located at RA=11:41:04.44, Dec=$-$26:35:08.6 (J2000.0) and the major axis of the region within the 3$\sigma$ contour is $\approx 3\farcm5$. The AKDE peak is slightly off-centre with respect to the 3 and 4 $\sigma$ isocontours; consequently for the rest of the paper we consider the overdensity region as a circle with 4\arcmin radius, centred at 1\farcm1 west/south-west from the AKDE peak, on RA=11:40:59.5, Dec=-26:35:11. This region is shown as a red circle in Figure~\ref{fig_over}. Note that the elliptical shape of the overdensity contours may be due to an edge effect; fewer sources are detected near the edge due to the drop in coverage.

\section{Surface density excess}
\label{sec:surf}

There are 76  SPIRE 250 \micron\ SXT sources at SNR above 3 and within the defined overdensity; 67 and 63 correspondingly at 350 and 500 \micron\ or $(1.51\pm0.17, 1.33\pm0.16, 1.25\pm0.16)$ arcmin$^{-2}$ in the three bands. In order to avoid  incompleteness bias for the faintest sources we only consider detections above our nominal $3\sigma$ flux limit of 20 mJy at 250 \micron, i.e. we have 43 detections or $0.86\pm0.13$ arcmin$^{-2}$. 

{In the four control fields, the surface densities of the 250 \micron\ detections above 20 mJy, derived using the area above 50\% relative coverage}, are $0.28\pm0.02$ arcmin$^{-2}$ (or $14\pm4$ sources rescaled to a 4\arcmin\ radius area) in COSMOS, $0.24\pm0.01$ arcmin$^{-2}$ (or $12\pm3$) in GOODS-North, $0.24\pm0.01$ arcmin$^{-2}$ (or $12\pm3$) in Lockman and $0.20\pm0.02$ arcmin$^{-2}$ (or $10\pm3$) in UDS. These correspond to $0.24\pm0.02$ arcmin$^{-2}$ on average, which means that there is a significant excess, a factor of 3.6, of the surface density in the overdensity with respect to the average levels in the control fields.

We also perform simulations of $4\arcmin$ regions in each one of the control fields, making sure that the whole area is within the 50\% relative coverage. From 1000 simulated areas the average numbers of 250 \micron\ detections above 20 mJy in the four control fields are $21\pm7$ in COSMOS, $21\pm 6$ in GOODS-North, $18\pm5$ in Lockman and UDS. The errors  are the st.dev. of the counts from the 1000 simulated regions. The interpretation of these results must be taken with caution because the simulated regions are correlated and the results are expected to be biased. Nevertheless, these results also support the significant excess of the surface density in the overdensity. 

\begin{figure*}
\centerline{
(a) \hfill \includegraphics[height=8cm]{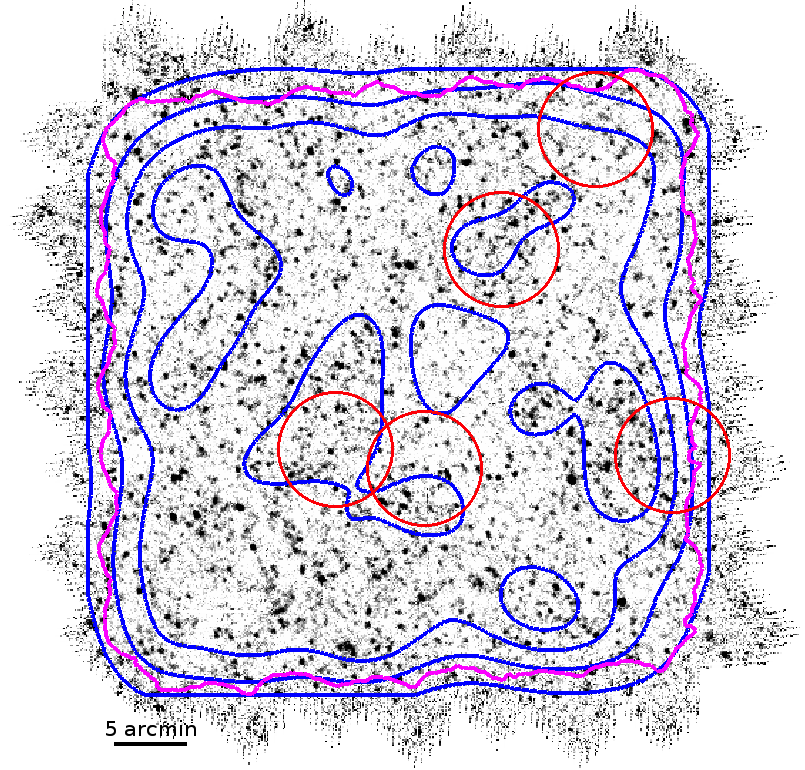} \hfill 
(b) \hfill \includegraphics[height=8cm]{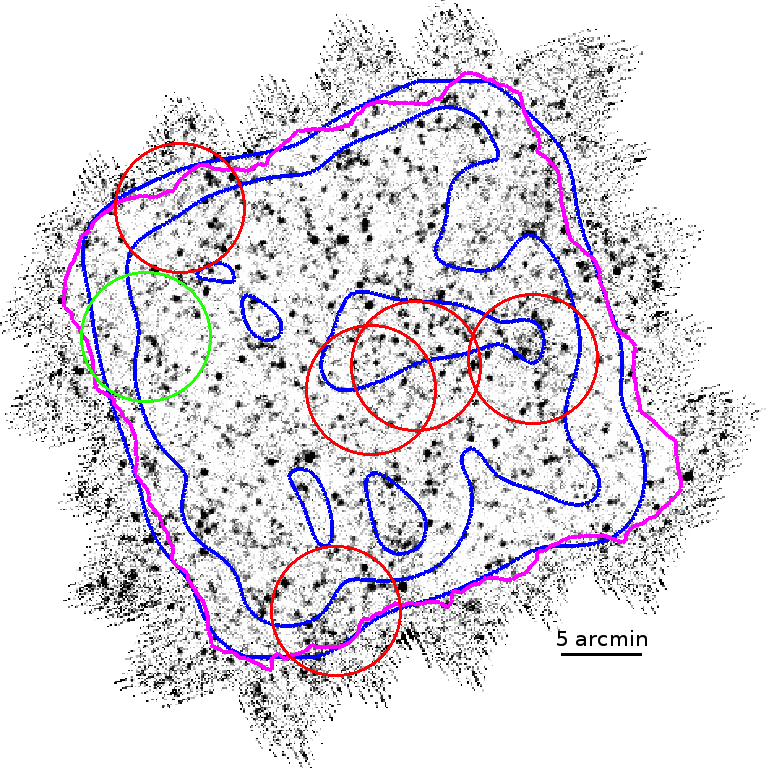}}
\centerline{
(c) \hfill \includegraphics[height=8cm]{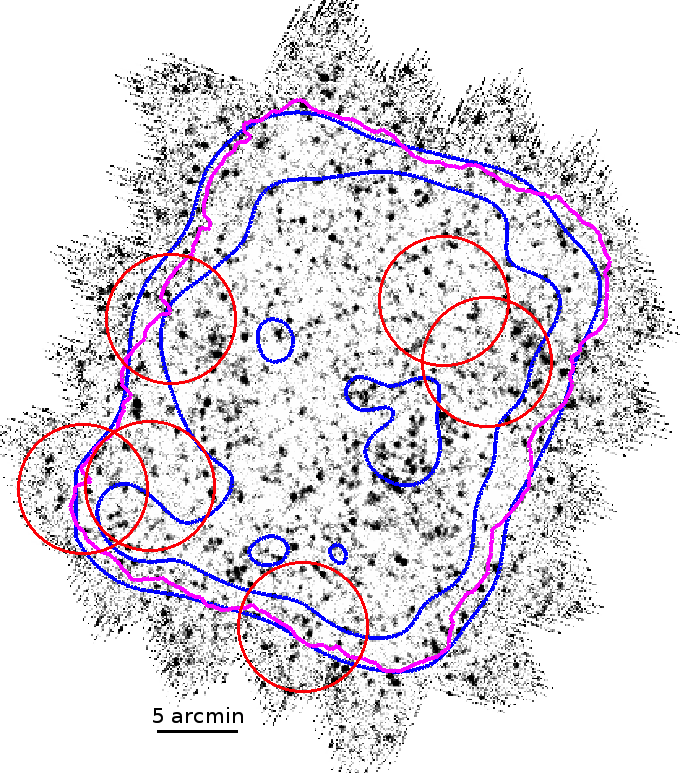} \hfill
(d) \hfill \includegraphics[height=8cm]{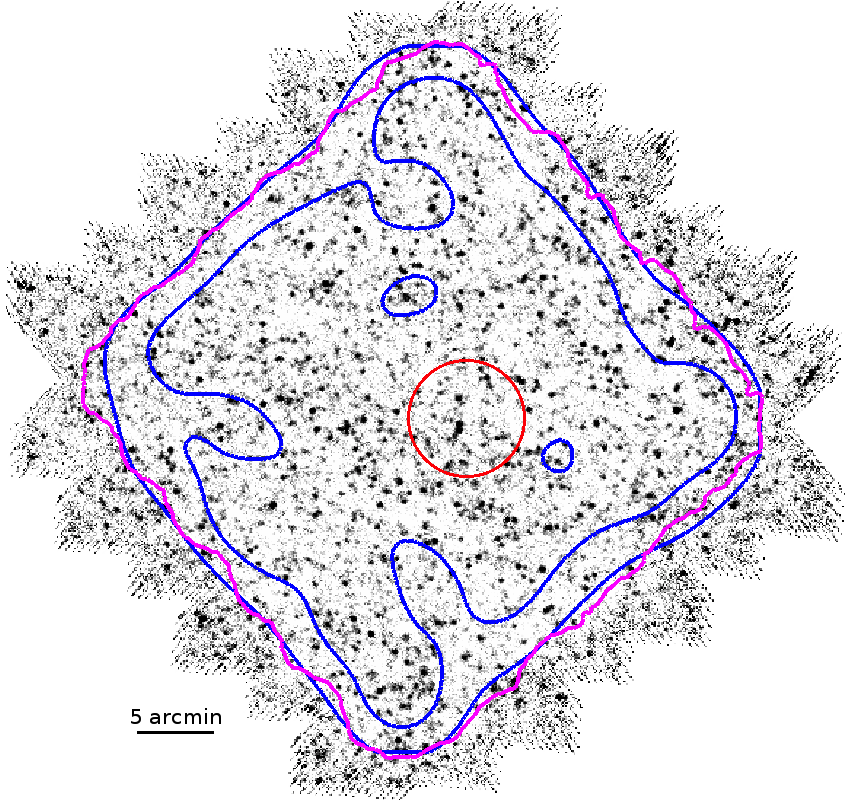}}
\caption{SPIRE 250 \micron\ maps in: \textbf{(a)} Lockman; \textbf{(b)} UDS; \textbf{(c)} COSMOS; and \textbf{(d)} GOODS-North. The data have been cut to match the depth of the Spiderweb map (see Section~\ref{sec_fields} for more details) and the greyscale ranges are the same in all maps, matching those in Figure~\ref{fig_over}. The AKDE contours run from 1 to 5 $\sigma$; the region above 50\% relative coverage is shown as a magenta contour. The highest peak in GOODS-North is at $\sim3\,\sigma$, $\sim 3.5\sigma$ for UDS and COSMOS and the highest one is at $\sim4.5\,\sigma$ in Lockman. The red circles of $4\arcmin$ radius show the positions of the photometrically selected candidate clusters at $1<z<1.6$ in Lockman, COSMOS and UDS from \citet{wen11}, keeping only those within the 50\% coverage. The only known confirmed protocluster in these fields, CLG\,0218 at $z=1.62$, is shown with a green circle in UDS \citep{papovich10}. In GOODS-North (d) we show the structure {candidate} at $z=1.99$, reported by \citet{chapman09}.}
\label{fig_fields}
\end{figure*}

The maximum number of detected sources from all $4\arcmin$ simulated regions are 42 in COSMOS, 35 in GOODS-North, 33 in UDS and 31 in Lockman.  So, there are regions of the same size and similar surface density in one (COSMOS) out of the three control fields.  It is interesting to note that the maximum AKDE in COSMOS is at $\sim3\sigma$ (see Figure~\ref{fig_fields}) which is significantly smaller than the $5.1\sigma$ we see in the overdensity. We interpret this discrepancy as being due to the more centrally concentrated distribution of the overdensity galaxies within the $4\arcmin$ region, as can be appreciated from Figure~\ref{fig_over}, where the overdensity region includes significant area with no sources outside the $3\sigma$ contour level. This effectively brings down the surface density.  This result shows that regions with similar surface density excess are rare but they can be found. However, they do not necessarily correspond to a peak in the spatial density distribution.

Note that without imposing a flux limit of 20 mJy at 250 \micron\ for the control fields we obtain  on average  $0.94\pm0.04$ arcmin$^{-2}$ -- this is close to the HerMES result {of 1.02 arcmin$^{-2}$} \citep{hermes} and makes the overdensity 1.51 arcmin$^{-2}$ only marginally in excess with respect to the average surface density in the control fields.

\section{SPIRE 250 \micron\ flux density distribution}
\label{sec:flux}

\begin{figure}
\centerline{
\includegraphics[width=8cm]{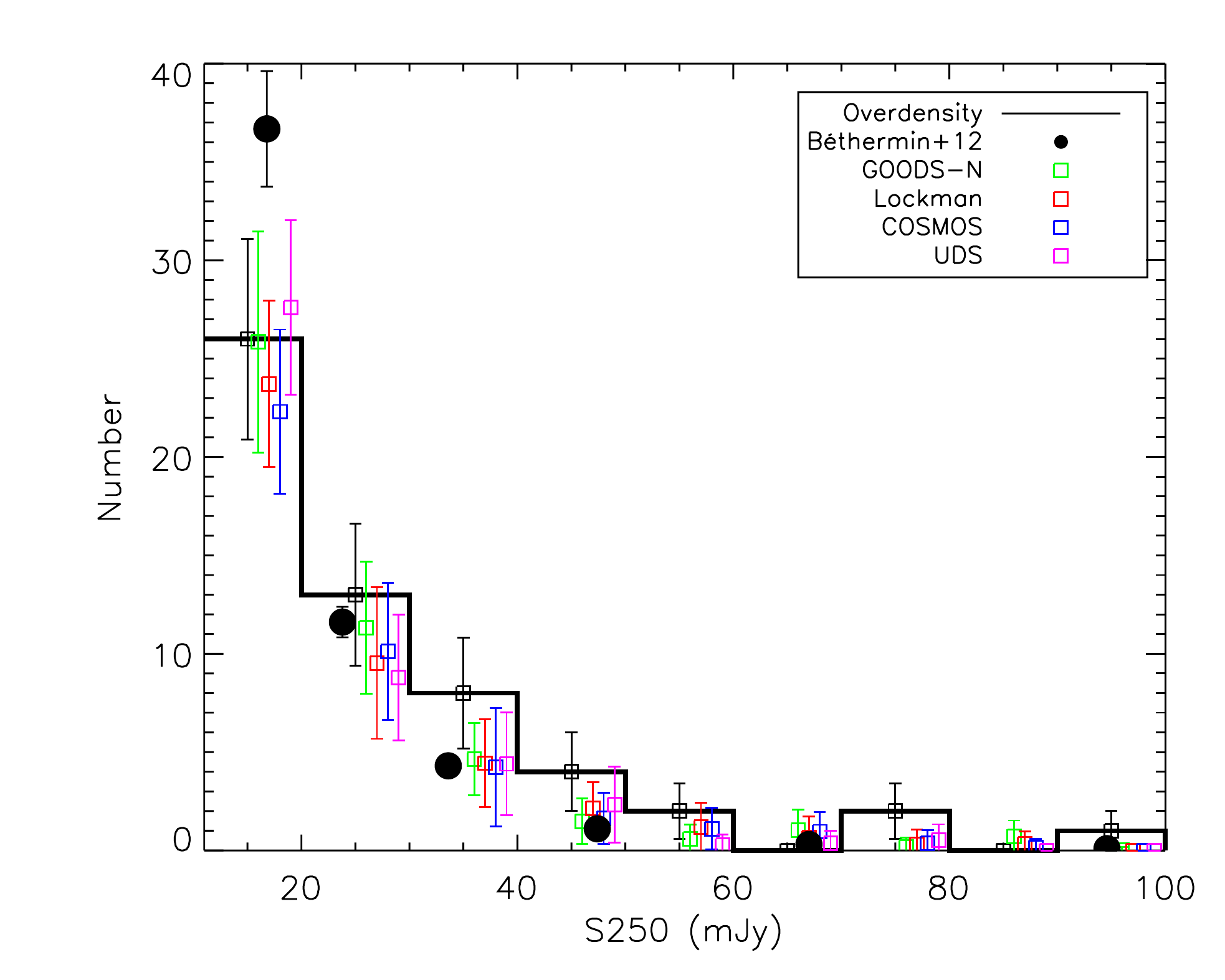}}
\caption{SPIRE 250 \micron\ flux density distribution for the overdensity (histogram and box symbol with error bar) and the four control fields (box symbols with error bars). The y-axis is the number of sources in a 4\arcmin\ area. Note that the control fields points were displaced from the centre of the bin for clarity. The black filled circles are the deep number counts from \citet{bethermin12}, rescaled to the area of the overdensity.}
\label{fig_s250}
\end{figure}

The 250 \micron\ flux density (S250) distribution of the sources in the overdense region is shown in Figure~\ref{fig_s250}. Using 1000 simulated $4\arcmin$ regions in the control fields, we derive the average 250 \micron\ flux distributions, which are also shown in Figure~\ref{fig_s250}, renormalised to the same number of sources as in the overdensity region. We see a marginal excess in the bin from 30 to 40 mJy: the expected number of sources in this bin is on average $4.4\pm2.0$ in a 4\arcmin\ radius, derived from the control fields. The deep number counts at 250 \micron\ derived from resolved sources in COSMOS from HerMES predict $4.3 \pm 0.3$ \citep{bethermin12}. We detect 8, which is a factor of almost two in excess, but at 1-2$\sigma$ only. The excess is quite likely to be partially due to the combined effect of source confusion and flux boosting, because of the closely packed sources in the clump. We also see a marginal excess in the 40-50 mJy bin. Nevertheless, the two-sided Kolmogorov-Smirnov (KS) test of the overdensity S250 distribution and each of the control fields indicate that they are significantly different in all cases: the probability, that the two distributions are drawn from the same parent one, is less than 3\%. 

\section{Far-infrared colours and modified black-body redshifts}
\label{sec:zbb}

\subsection{Ancillary data}
\label{sec_aux}

There are no ground-based optical or NIR observations nor \textit{Spitzer} or \herschel-PACS observations that cover this region  $\sim7\arcmin$ south of the Spiderweb radio galaxy (details on the available multi-wavelength data in this field are presented in \citealt{koyama13}). Hence we utilise the available all-sky surveys, of which \textit{WISE} \citep{wise} is the most useful. In general, the depth of the \textit{WISE} band W1 allows us to detect an $L_{\star}$ galaxy at $5\,\sigma$ out to $z\approx 1$ \citep{gettings12}. This is rather shallow for SPIRE detected galaxies, which are predominantly at redshifts greater than 1. Nevertheless, these additional data are useful to identify low redshift sources. There are 158 WISE detected galaxies in the overdense region of $\sim 4\arcmin$ radius using the All Sky Data Release from March 2012. From those, 21 are within $5\arcsec$ of a SPIRE detected galaxy and have a $>3\,\sigma$ detection in band W1 (3.4 \micron). We do not use the \textit{WISE} photometry for these 21 sources in any of the following analysis, but we simply flag the SPIRE sources that may be at low redshift ($z<1$).

\subsection{Modified black-body redshift}

It is a challenge to derive good photometric redshifts when having only three bands (250, 350 and 500 \micron) at best, each with a flux calibration uncertainty of $\sim7$\% \citep{swinyard10,bendo13}. However, there are already a number of  \herschel-based studies applying different techniques in order to overcome the complications of large beams, flux boosting and substantial flux uncertainties, especially for faint, distant, star-forming galaxies (e.g. \citealt{lapi11, roseboom12, casey12a, chakrabarti12}). Hence, with no supporting ancillary data we decide to approach the problem using a simple fit with a redshifted, single-temperature, modified black-body:
\begin{equation}
S_\nu(z) \propto \nu_z^{\beta}\,\mathrm{B}(T,\nu_z),
\label{eq_bb}
\end{equation}
where $\nu_z = \nu/(1+z)$ is the redshifted frequency, $\beta$ is the dust emissivity spectral index, which we keep fixed at 1.5 (e.g. \citealt{magnelli12}) and $\mathrm{B}(T,\nu_z) = \frac{2h\nu_z^3}{c^2}\,\frac{1}{e^{h\nu_z/kT}-1}$ is the Planck function. With at most 3 photometric detections, we proceed by fixing the temperature $T$ and leaving the redshift $z$ as a free parameter. For \herschel\ detected star-forming galaxies the representative dust temperatures are between $\sim 20$ K and $\sim 60$ K (e.g. \citealt{magdis10,magnelli12}). Thus, we fix the temperature to $30$ K, which is also close to the dust temperatures of 30-34 K of the four confirmed Spiderweb cluster members with spectroscopic redshifts (see later and S\'anchez-Portal, in preparation). We further constrain the fit to sources that have at least 2 out of 3 photometry detections, which brings down the number of sources used in the $z_\mathrm{BB}$ analysis to 58. Even with the simplification of fixing $T$ and $\beta$, the $\chi^2$ statistic is not very useful because we have at most 2 degrees of freedom and we fit two parameters: redshift and overall normalisation. This means that the error on the parameters is not representative. 

We call the best-fit redshift the black-body redshift and denote it $z_\mathrm{BB}$. Galaxies with similar far-infrared (FIR) SEDs and colours would have similar $z_\mathrm{BB}$.  We note that there is a strong degeneracy between $T$ and $z_\mathrm{BB}$\footnote{From the Wien's displacement law, $\lambda_0\,T = \lambda_{\mathrm{obs}}\,T/(1+z) = const.$, where $\,z$ is the redshift,  $\lambda_0$ is the wavelength of the black-body peak in rest-frame and $\lambda_{\mathrm{obs}}$ is the peak location in observed wavelength. Consequently, for the same black-body SED, $T/(1+z)$ is constant, see also \citet{blain03}}, and by definition, $z_\mathrm{BB}$ would match the actual redshift only in some particular situations, i.e. dust temperature and dust emissivity $\beta$ close to our fixed values.

We compare the derived $z_\mathrm{BB}$ for the \herschel-SPIRE detected galaxies with spectroscopic redshifts in \citet[][hereafter C12]{casey12b}. As expected, the Pearson correlation coefficient between the two is only 0.2, which means that there is practically no correlation between the spectroscopic redshift and the derived $z_\mathrm{BB}$ at $z<2$ (the limit of the spectroscopic data in C12). Nevertheless, if there is an excess of galaxies with similar colours (i.e. at a similar redshift) then the distribution of $z_\mathrm{BB}$ would be informative and would show this excess. 

\begin{figure}
\centerline{
\includegraphics[width=9cm]{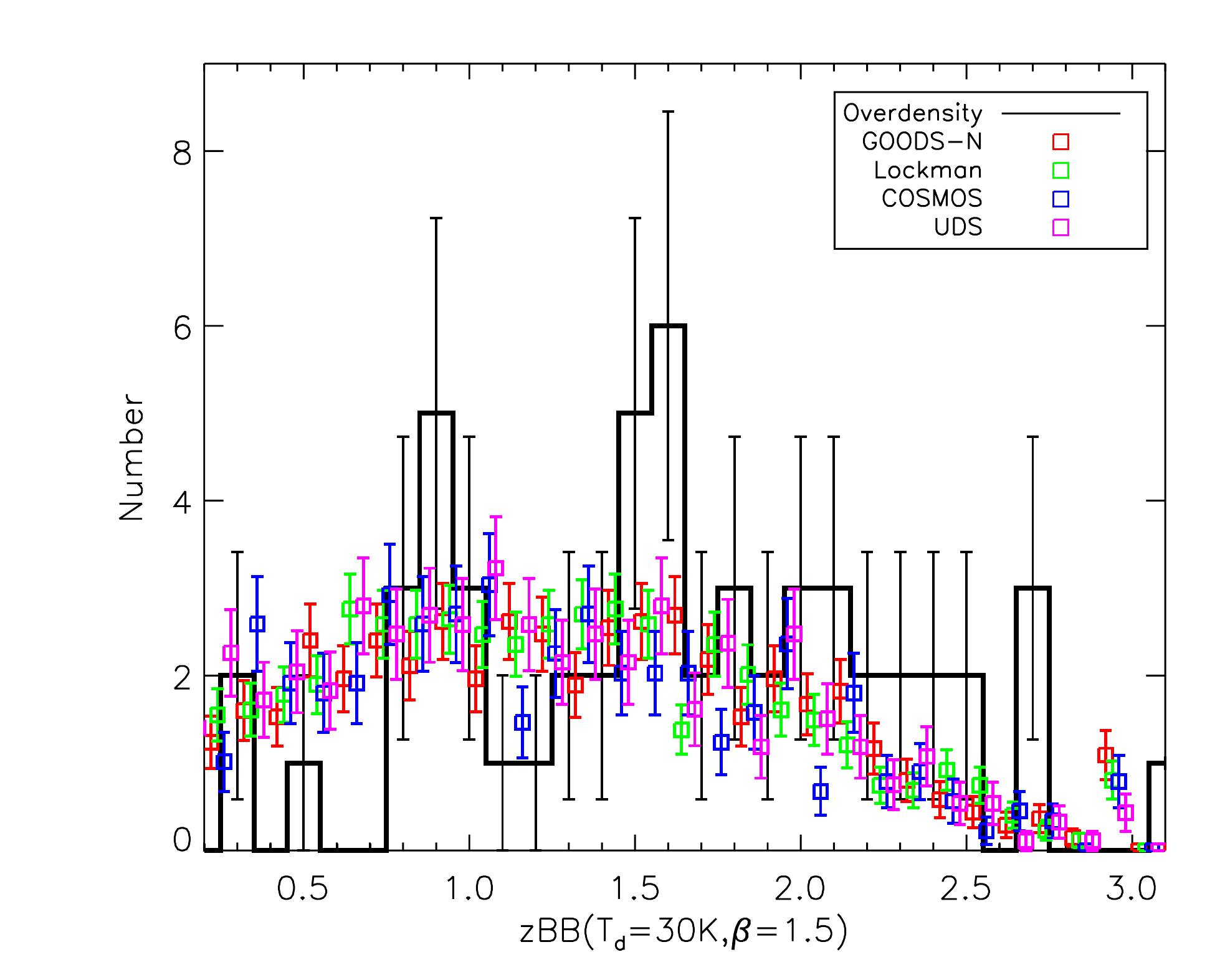}
}
\caption{Distribution of the FIR redshifts $z_\mathrm{BB}$ for the overdensity region (histogram) and the four control fields (box symbols with error bars, symbols displaced within the bins for clarity). The bin size is 0.1 and all error bars are Poissonian. The distributions in the fields were renormalised to the same number of sources as in the overdensity.}
\label{fig_tdust}
\end{figure}

Figure~\ref{fig_tdust} shows the derived $z_\mathrm{BB}$ distribution in the overdensity and we note that indeed there is a significant peak at $z_\mathrm{BB} \sim 1.5$. Following the same approach, calculating $z_\mathrm{BB}$ for the four control fields, we see that the distributions are similar to each other, while they are significantly different from the overdense field (see Figure~\ref{fig_tdust}), especially for the two $z_\mathrm{BB}$ bins centred on 1.5 and 1.6, although there are some hints of excess at $z_\mathrm{BB} \approx 0.9$ and $\approx 2.6$ too. The two-sided KS test gives the maximum probability of $<$5\% that the overdensity and the fields $z_\mathrm{BB}$ distributions are drawn from the same parent distribution. Using the average level and Poissonian error of $2.4 \pm 0.5$ at $z_\mathrm{BB}=1.5$ in the fields, we estimate the peak signal-to-noise ratio of more than 7.  We must emphasise that the location of the $z_\mathrm{BB}$ peak depends strongly on the assumed dust temperature and could range from 2.33 ($T=40$ K) to 0.7 ($T=20$ K). The effect of $\beta$ on $z_\mathrm{BB}$ is not significant, i.e. changing $\beta$ from 1.5 to 2.0, for $T=30 K$, will only change $z_\mathrm{BB}$ by less than 5\%.

We also check if the significant $z_\mathrm{BB}$ peak is not narrower than expected for a natural spread of the galaxies SEDs and flux measurements. We use the empirical SED templates from \citet[][hereafter K13]{kirkpatrick13}, derived from 24 \micron\ detected galaxies that were observed with \textit{Herschel}. We select the SED template for a $z=2$ star-forming galaxy and redshifted it to $z=1.5$, normalising the 250 \micron\ flux density to 30 mJy (where we see excess sources, see Sect.~\ref{sec:flux}). We generate flux densities at 250, 350 and 500 \micron\ randomly from within the SED template flux confidence range, incorporating the 7\% flux calibration uncertainty and the confusion noise of 6 mJy in quadrature for the measurement error. Then we fit a modified black-body for the $z_\mathrm{BB}$, following exactly the same procedure as for the overdensity sources. From 1000 simulated galaxies we obtain $z_\mathrm{BB} = 1.50 \pm  0.31$, where the spread $\Delta z_\mathrm{BB}$ of 0.31 is obtained as the st.dev. of all $z_\mathrm{BB}$ measurements.  Similar results are obtained using the other SED template in K13 -- for star-forming galaxies at $z=1$. This result confirms the correct approach on estimating $z_\mathrm{BB}$, as we recover exactly the simulated redshift and also provide us with the expected spread of the measurements $\Delta z_\mathrm{BB}$. This $\Delta z_\mathrm{BB}$ of 0.31 is somewhat larger than the observed one with width of 0.2 for the two significant bins, which may be an indication that the spread in the SED shapes in the galaxies within the peak is smaller than the spread in the empirical template in K12.

\begin{figure}
\centerline{
\includegraphics[width=9cm]{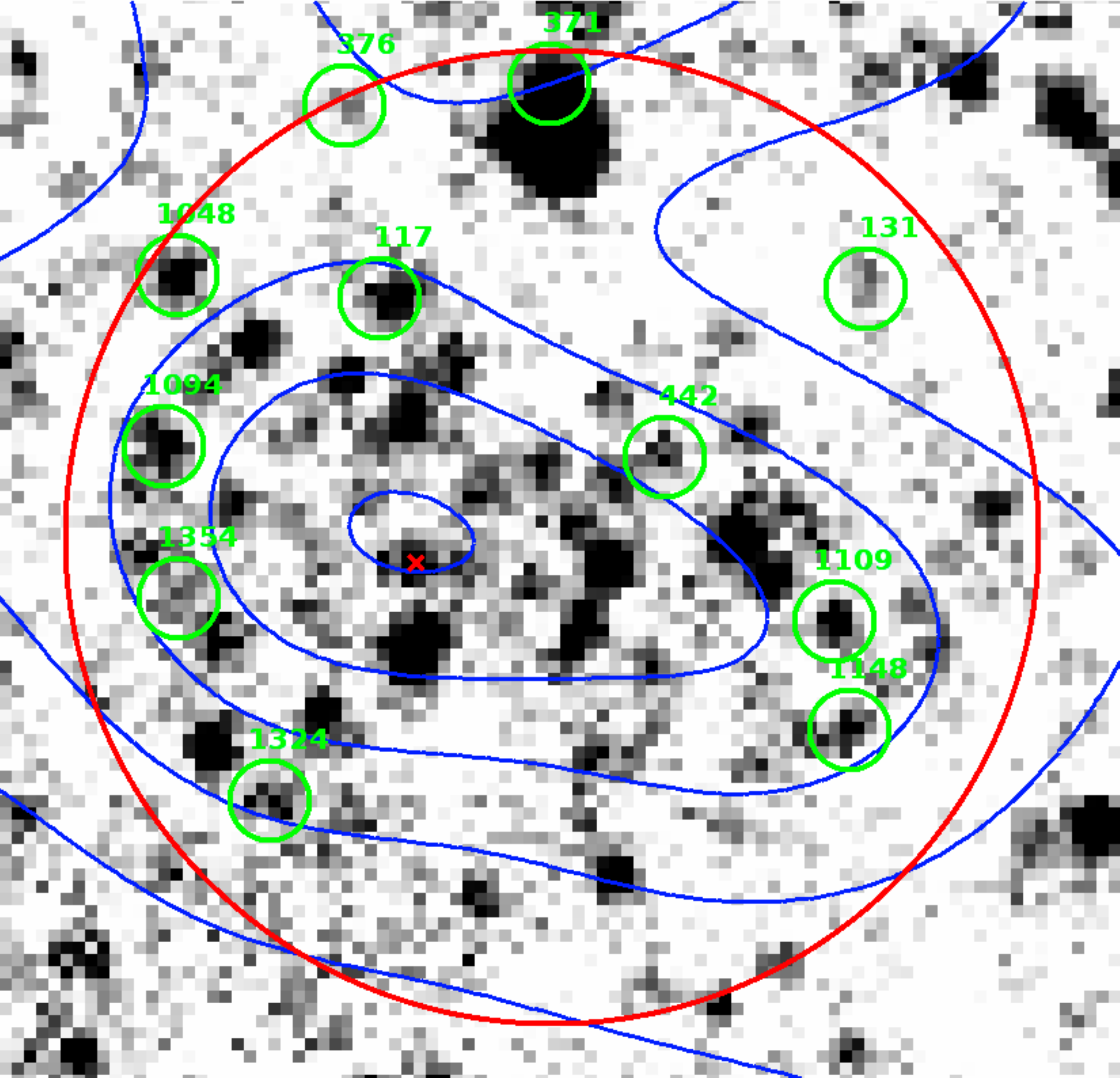}
}
\caption{Zoom of the overdensity region showing the locations of the 11 sources with $z_\mathrm{BB}$ within [1.45,1.65]. The 250 \micron\ image grey colour scale cuts, the AKDE contours, the circle and the '$\times$' symbol are the same as in Figure~\ref{fig_over}.}
\label{fig_zoom}
\end{figure}

There are 11 sources within the peak ($1.45 \leq z_\mathrm{BB} \leq 1.65$). Figure~\ref{fig_zoom} shows their IDs {from the SXT catalogue} and their distribution within the overdensity: they are mostly between the 3 and the 4$\sigma$ AKDE contours. It is interesting to note that they are not {preferentially found near the AKDE peak}. Nevertheless, this is a significant excess of sources with similar $z_\mathrm{BB}$ and thus similar colours and SED shapes. 

\begin{figure*}
\centerline{
\includegraphics[width=4cm]{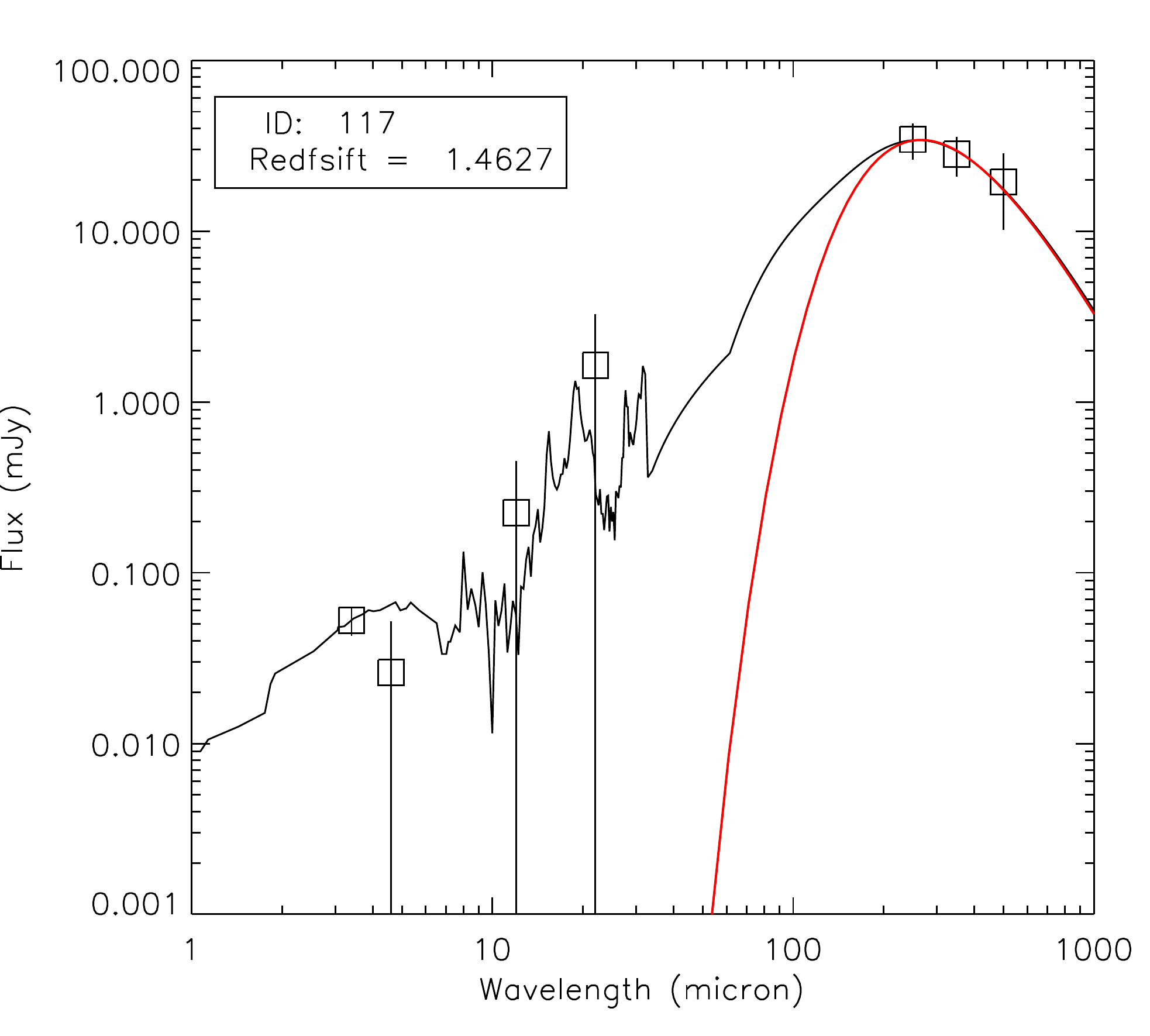}
\includegraphics[width=4cm]{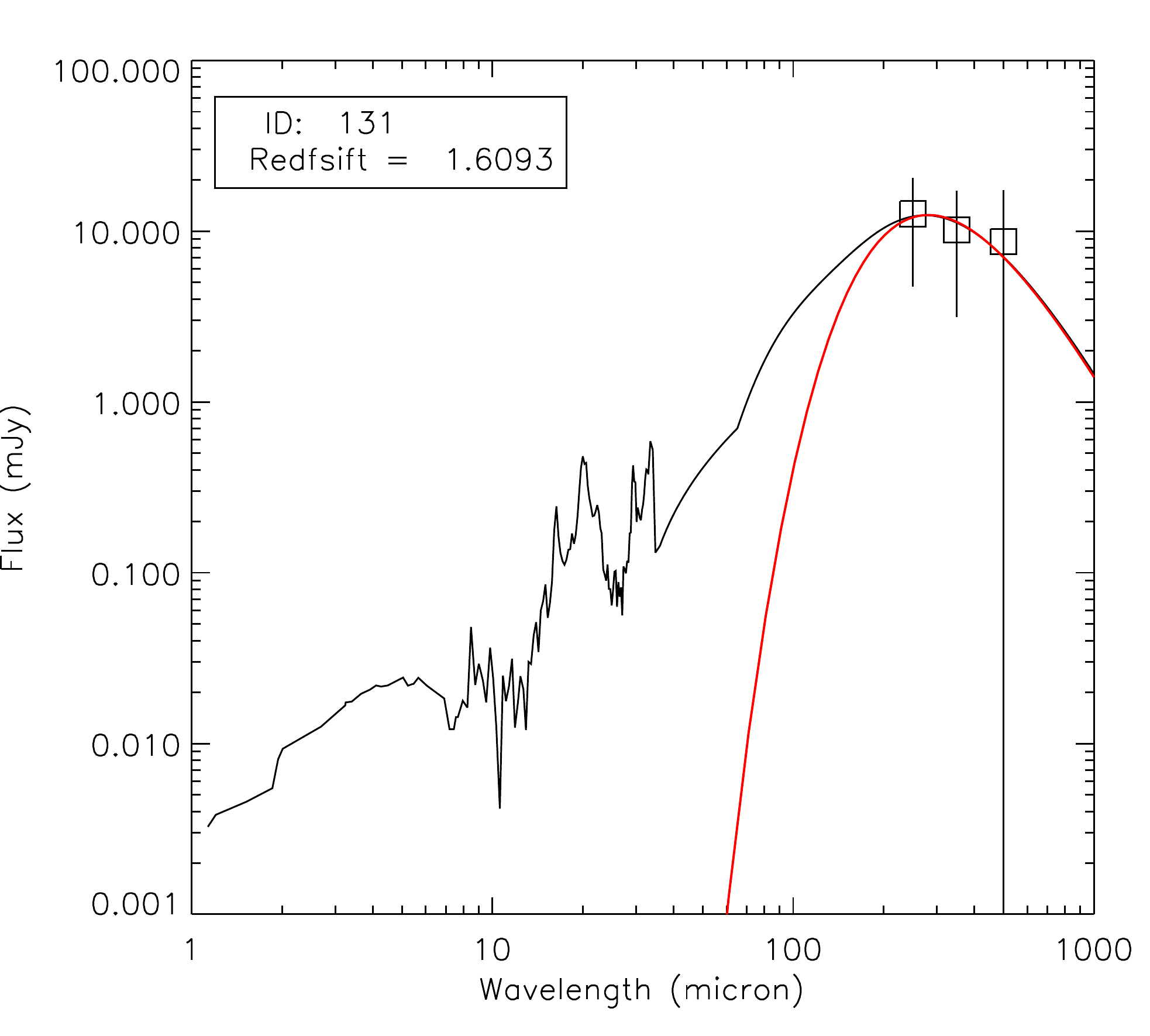}
\includegraphics[width=4cm]{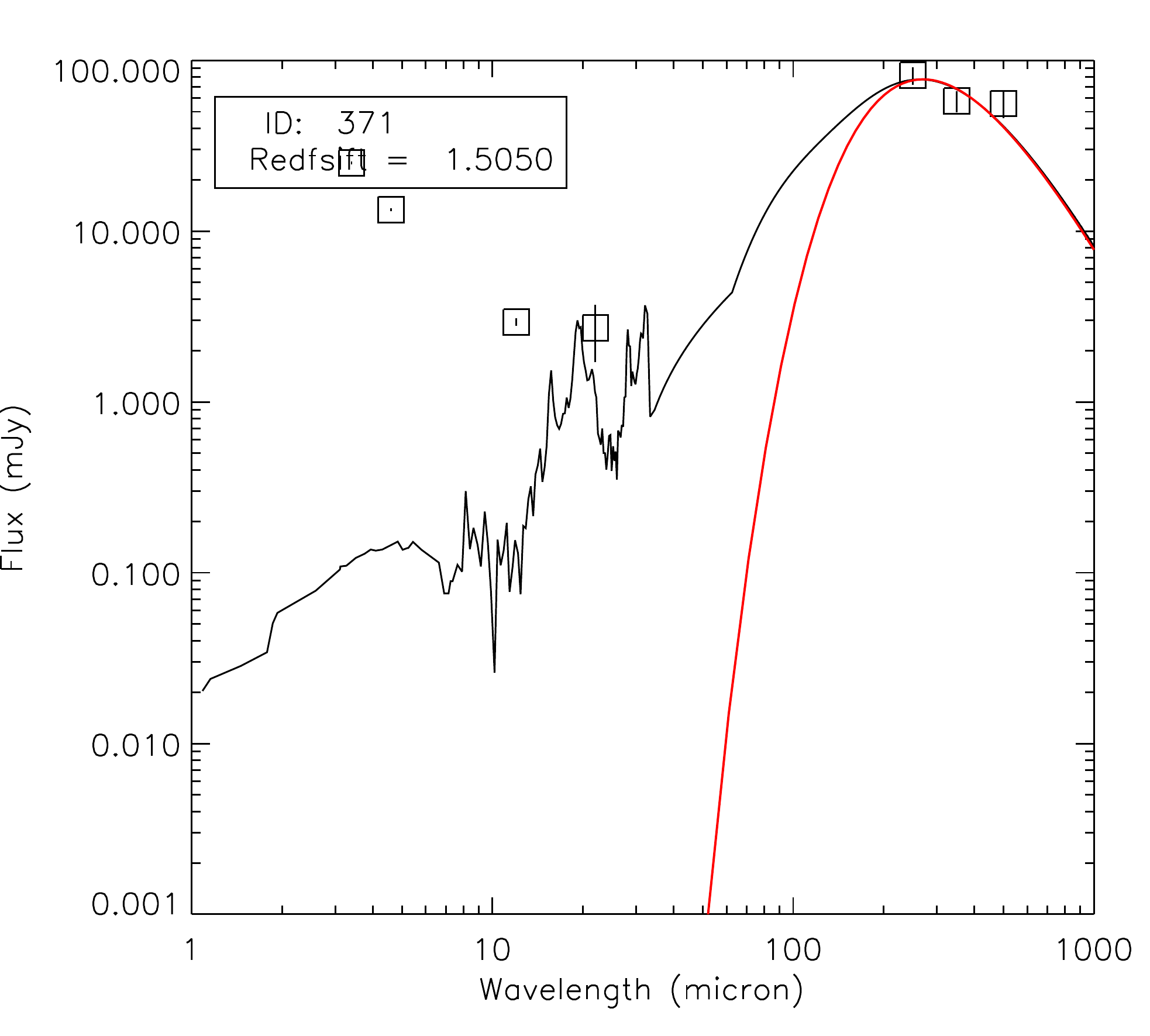}
\includegraphics[width=4cm]{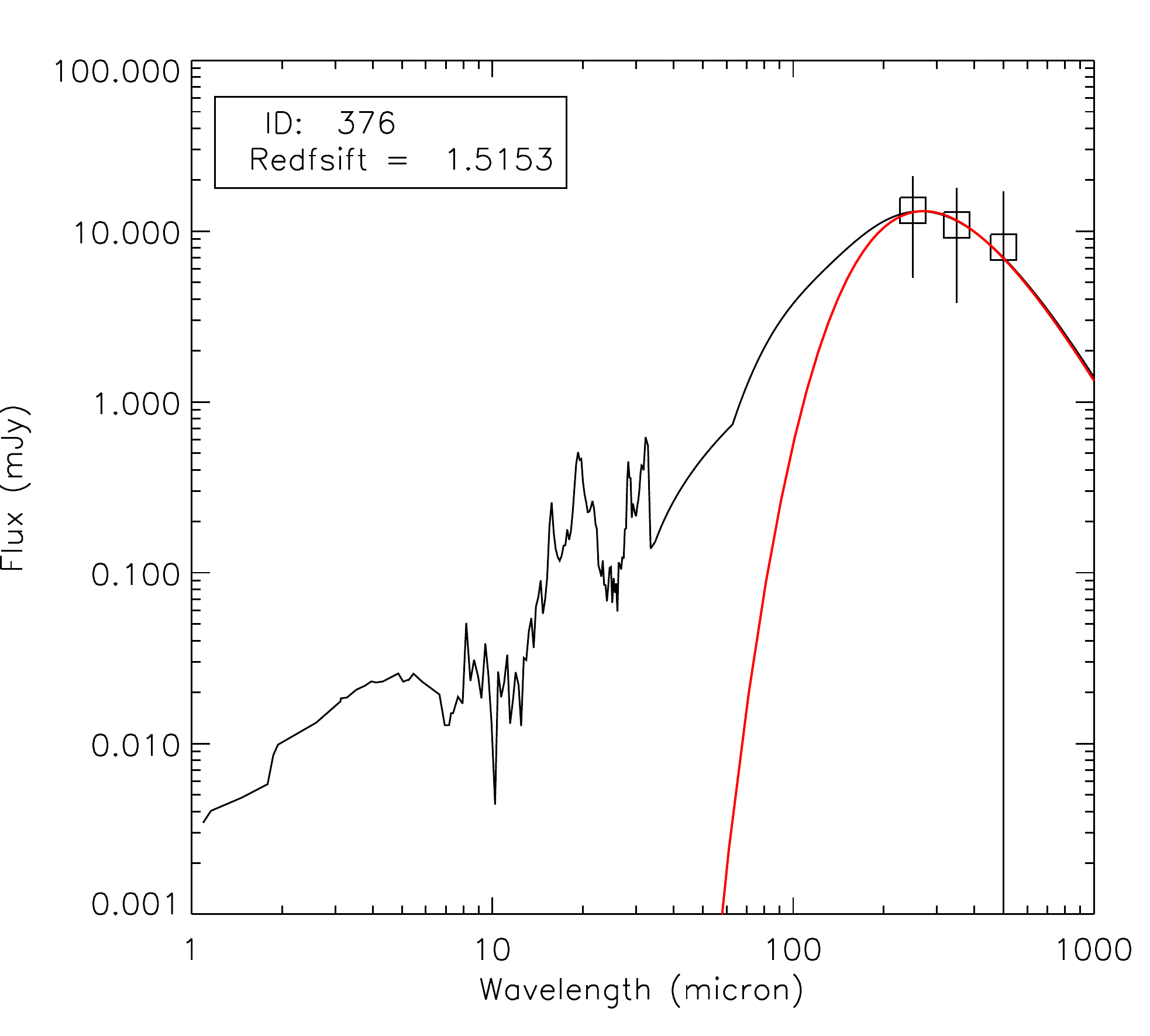}}
\centerline{
\includegraphics[width=4cm]{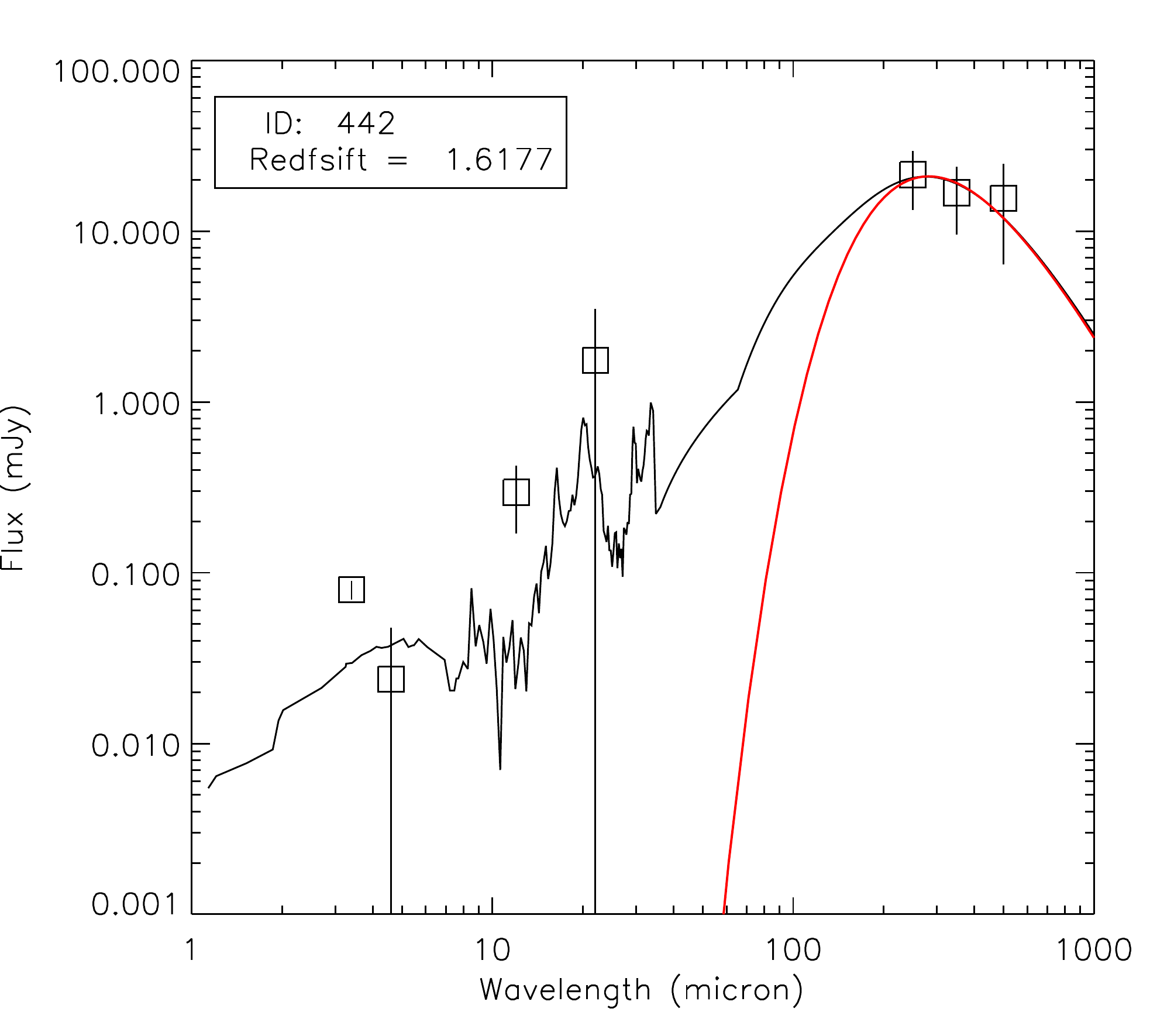}
\includegraphics[width=4cm]{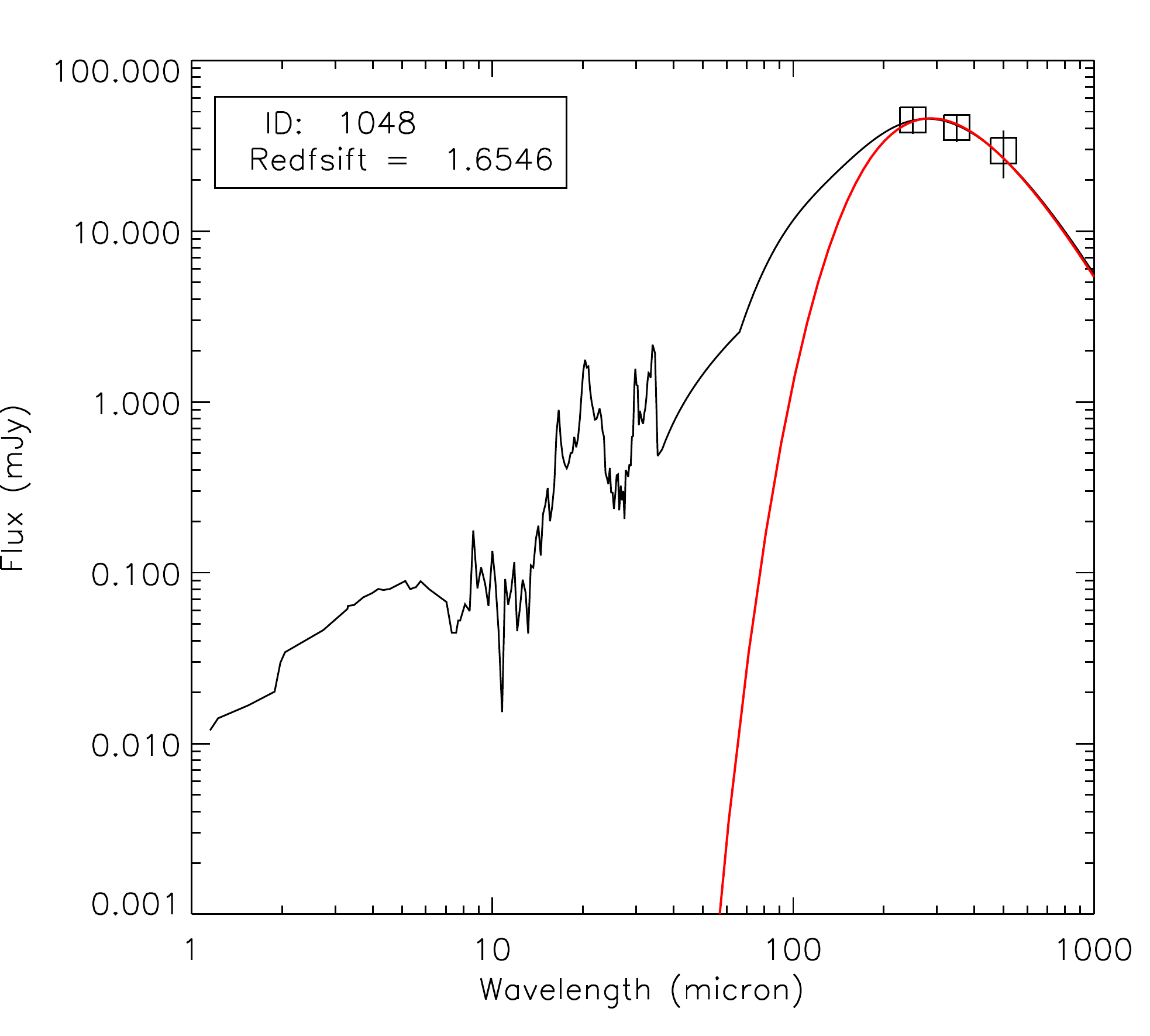}
\includegraphics[width=4cm]{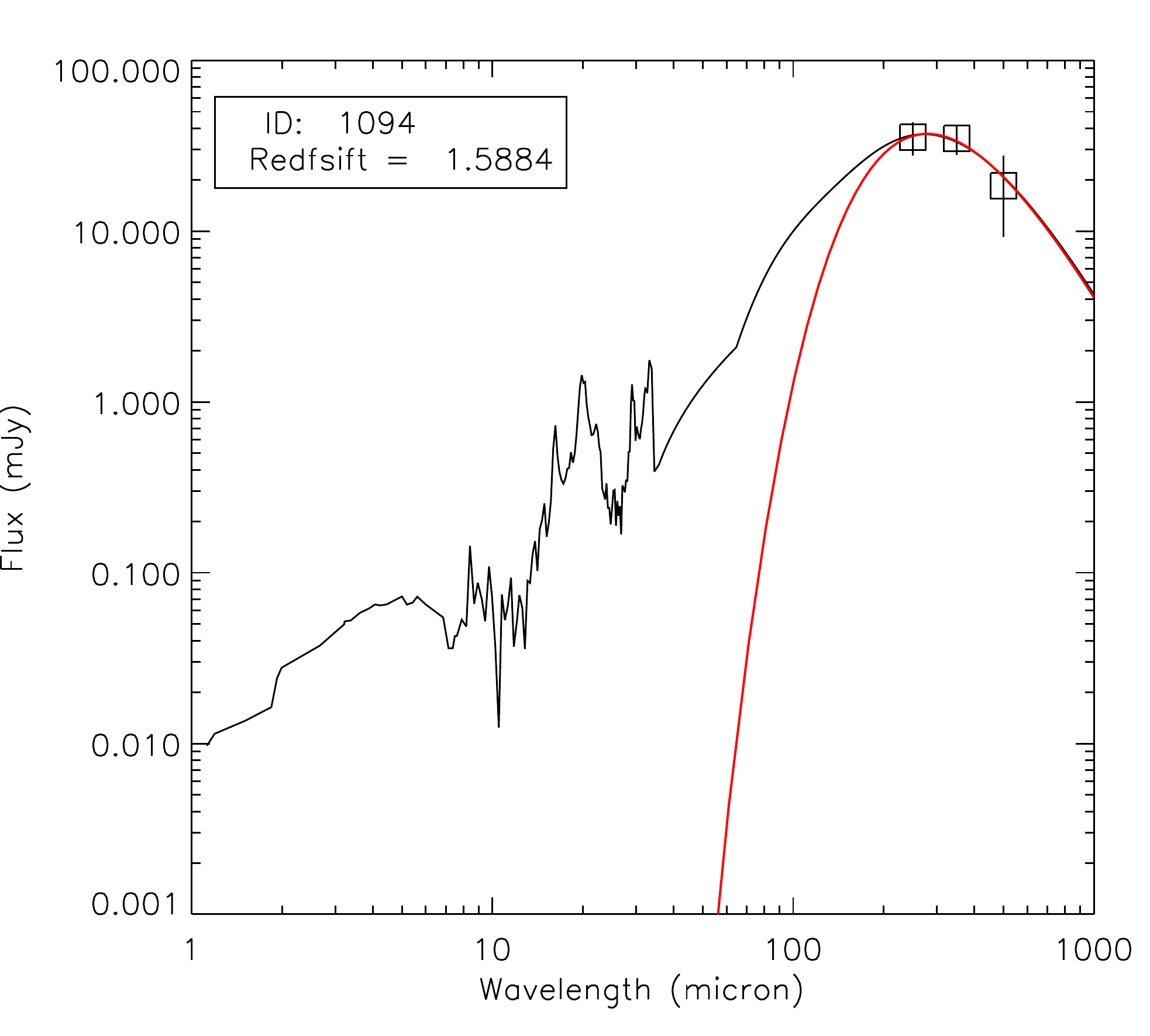}
\includegraphics[width=4cm]{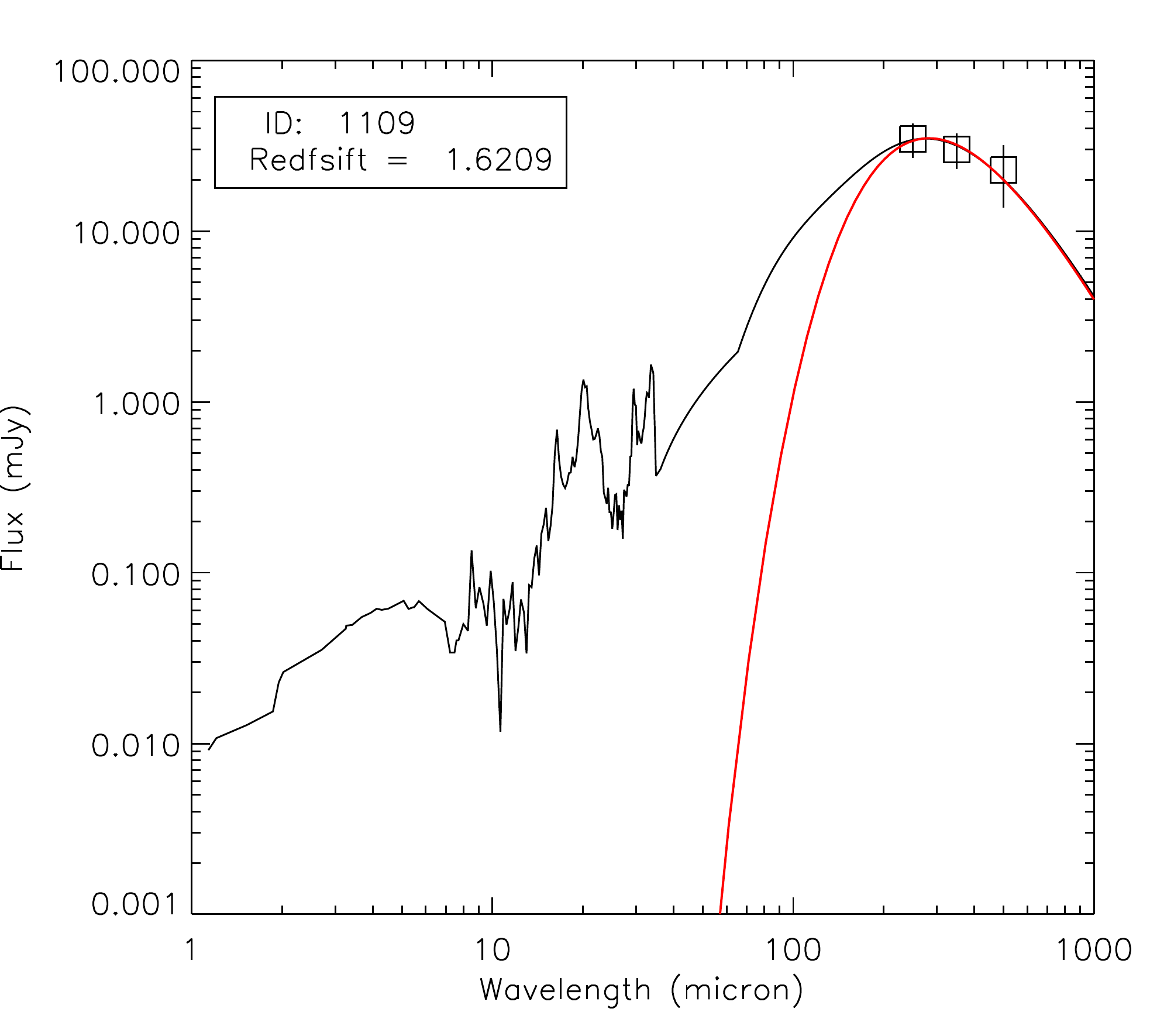}}
\centerline{
\includegraphics[width=4cm]{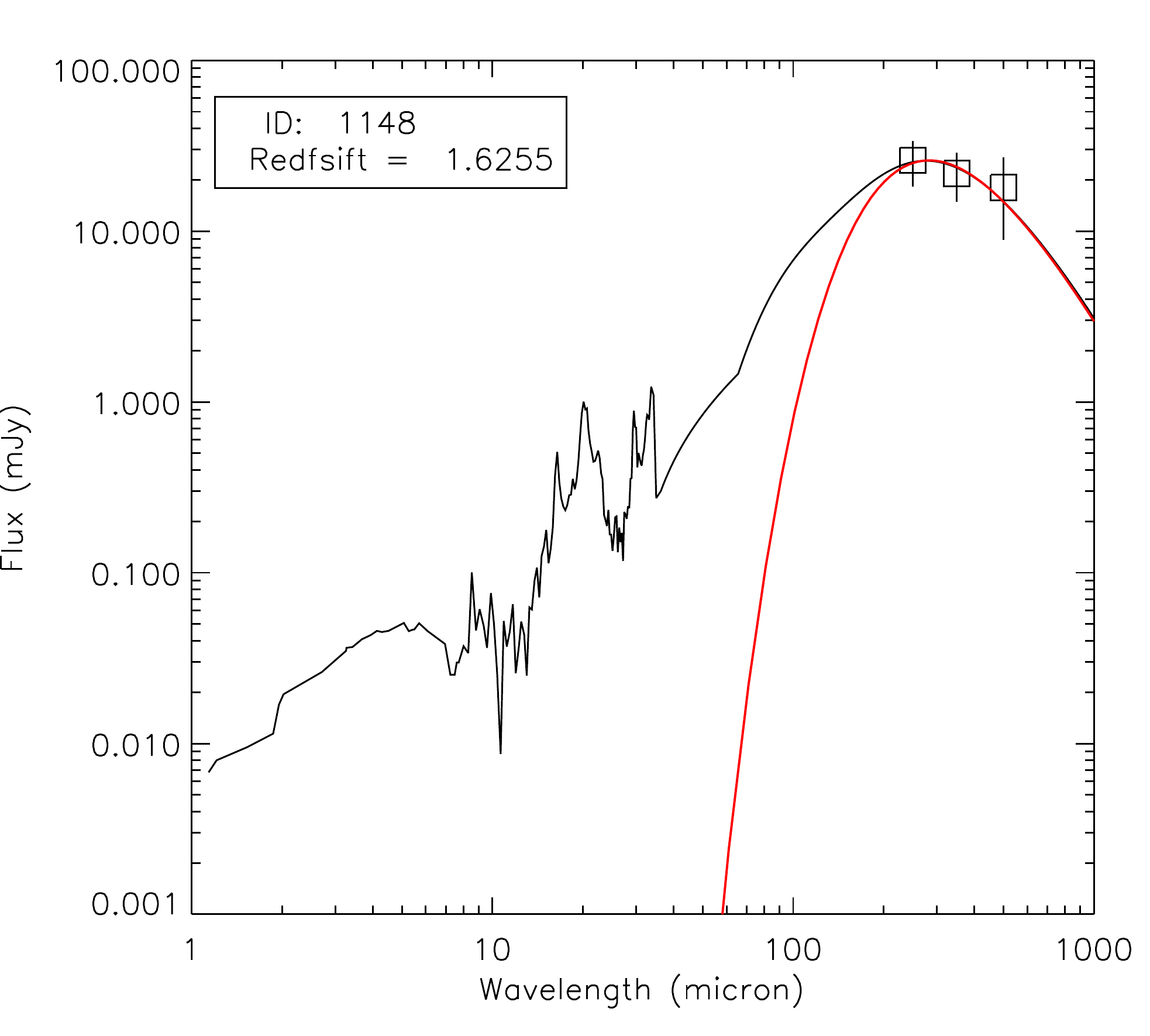}
\includegraphics[width=4cm]{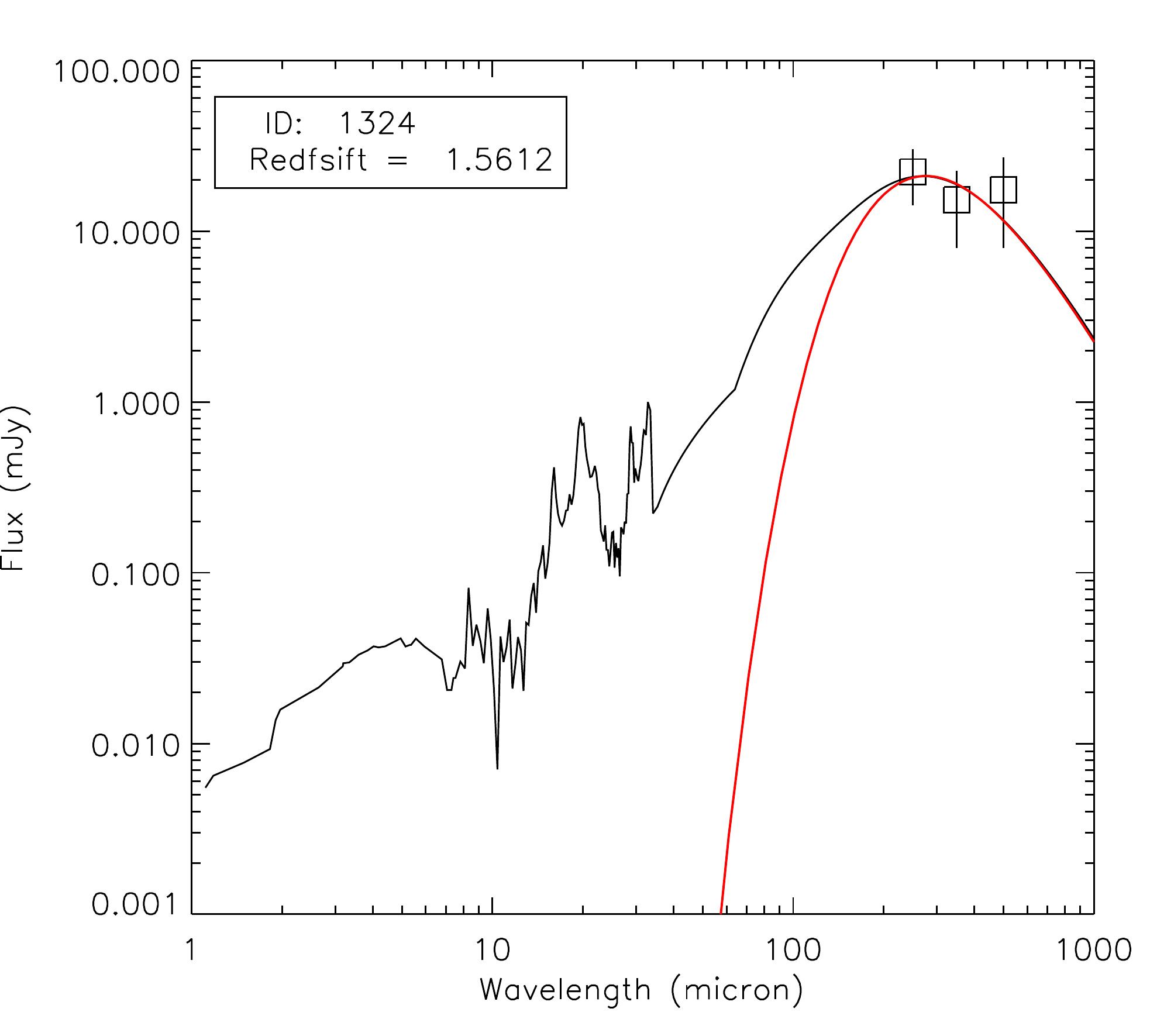}
\includegraphics[width=4cm]{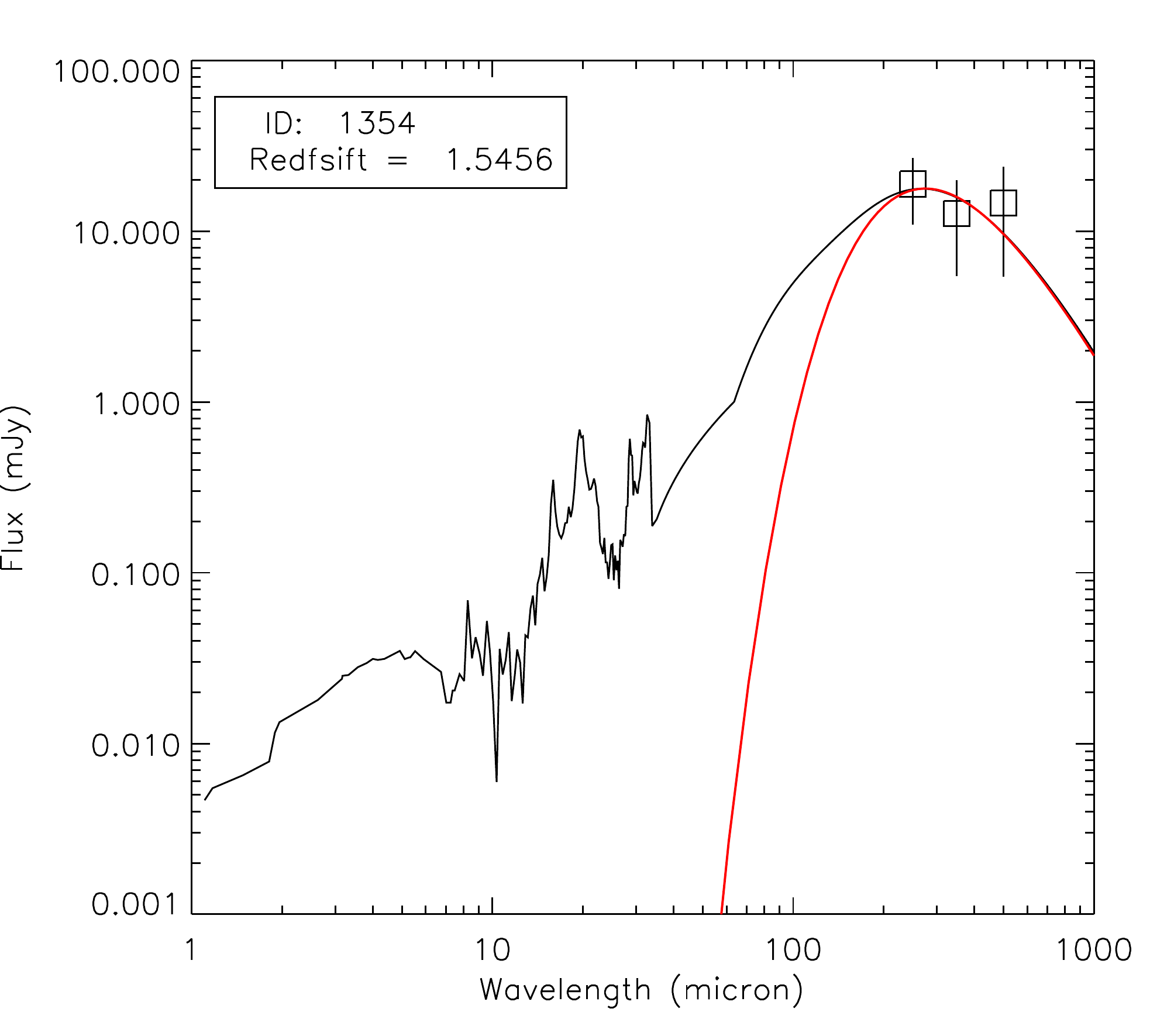}
\hspace{4cm}}
\caption{{SED and black-body fits for the 11 sources} in the $z_\mathrm{BB}$ peak. The best blackbody fit is shown in red.  The legend provides the IDs and the derived $z_\mathrm{BB}$. The SED template (in black) is an empirical SED of a star-forming galaxy at $z=2$ from K13. {The SED template was redshifted to $z_\mathrm{BB}$ and we fit only for the normalisation.} The photometric upper limits are shown with error bars extending down to the y-axis lower limit, three sources has WISE detections or upper limits (see Section~\ref{sec_aux}).}
\label{fig_zbbfit}
\end{figure*}

Figure~\ref{fig_zbbfit} shows the modified black-body fit (as a red curve) for these 11 sources as well as the SED template from K13, fixed to the $z_\mathrm{BB}$ redshift.  {Of the 11 sources, 3 have WISE W1 (3.4\micron) counterparts above the photometric limits: IDs 117, 371 and 442; one of them (ID\,371) is detected in W2 (4.6 \micron) and one (ID\,442) in W3 (12\micron)}. The photometry of ID\,371 is likely affected by the neighbouring very bright 250 \micron\ source and its detection by WISE at 3.4 and 4.6 \micron\ is probably indicative of low redshift or a blended source. 

\subsection{Colour-magnitude diagram}

Figure~\ref{fig_cmag} shows S250 vs S250/S350 colour-magnitude diagram of the overdensity sources, compared to the colour-magnitude results from the spectroscopic sample in C12 for $z<1$ and $z>1$ galaxies. The two regions occupied by the C12 galaxies were constructed by keeping only sources with S250/S350 $< 3$ and S250 $< 100$ mJy (thus removing some outliers with peculiar colours or higher fluxes). We already demonstrated that there is no correlation between the C12 spectroscopic redshifts and the black-body redshifts, and Figure~\ref{fig_cmag} shows it clearly: the two samples of C12 galaxies at $z<1$ and $z>1$ largely overlap and render the S250/S350 colour information nearly useless as a crude redshift estimator. We note, however, that the sources in the $z_\mathrm{BB}$ peak are within a small range [1.0,1.5] of the S250/S350 colours (indicated by dashed line in Figure~\ref{fig_cmag}), which is in support of our use of $z_\mathrm{BB}$ as a simple way to identify galaxies with similar observed SEDs (i.e. some combination of temperature and redshift).

\subsection{Connection with the Spiderweb protocluster?}

\begin{figure}
\centerline{
\includegraphics[width=8cm]{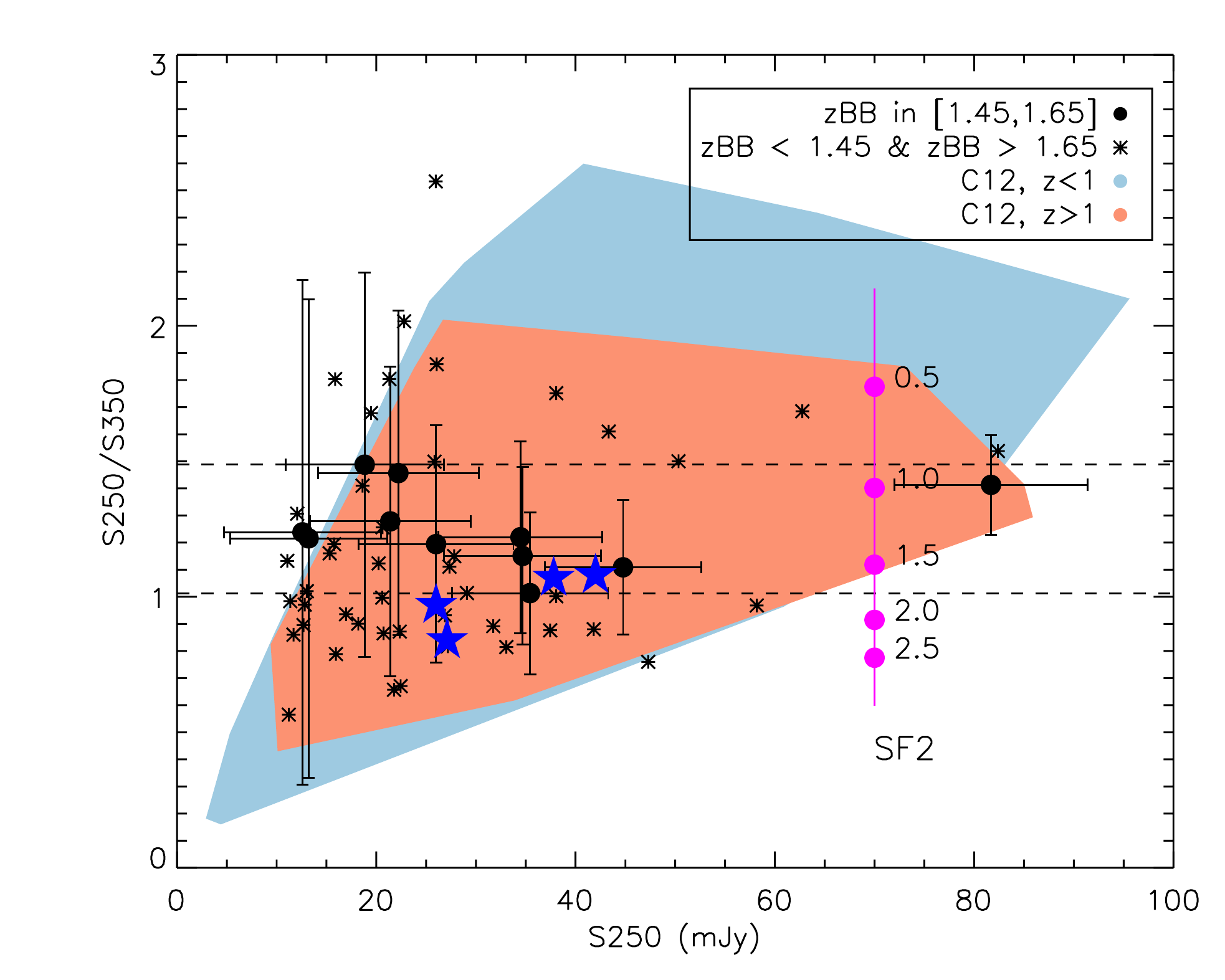}
}
\caption{The S250 vs S250/S350 colour-magnitude diagram for the overdensity galaxies. The filled polygons highlight the regions occupied by the  \herschel\ detected galaxies with spectroscopic redshifts in C12: blue for $z \leq 1$ and red for $z>1$. Note that the two regions overlap. The S250/S350 track for the SF2 SED (from K13) is shown as a vertical magenta line with the redshifts indicated. We only show the error bars for the 11 overdensity sources with $1.45 \leq z_\mathrm{BB} \leq 1.65$. Two dashed lines mark the upper and lower limits in S250/S350 colour for the galaxies with S250 above 20 mJy ($3\times\,rms$) in the $z_\mathrm{BB}$ peak. The four SPIRE-detected counterparts to spectroscopically confirmed Spiderweb galaxies (S\'anchez-Portal, in preparation) are shown as filled blue stars.}
\label{fig_cmag}
\end{figure}

It is interesting to note that the overdensity can in principle be part of a large complex at the redshift of the Spiderweb protocluster. As reported in \citet{koyama13}, there is a large-scale filament in the Spiderweb field traced by H$\alpha$ emitters of a comoving size of $\gtrsim10$ Mpc, going from NW to SE. Unfortunately, the narrow band filter observations do not reach as far as the overdensity region. A preliminary analysis of the \herschel\ data identify only four SPIRE-detected counterparts of spectroscopically confirmed Spiderweb protocluster galaxies (S\'anchez-Portal, in preparation), including the radio galaxy \citep{seymour12}. Their S250/S350 colours, as shown in Figure~\ref{fig_cmag}, are similar to the overdensity galaxies in the $z_\mathrm{BB}$ peak. In addition, due to the temperature-redshift degeneracy, the peak of $z_\mathrm{BB}$ would match the protocluster redshift at $z=2.2$ if we fix the temperature to 38 K instead of 30 K. We do not want to enter into further speculation regarding this scientifically interesting hypothesis, because there are large uncertainties in the colour-magnitude diagram and there are too few  SPIRE-detected confirmed Spiderweb galaxies. We presently lack auxiliary data that may help to establish the veracity of this claim.

\section{Planck view of the overdensity}
\label{sec:planck}

\begin{figure}
\centerline{
\includegraphics[width=8cm]{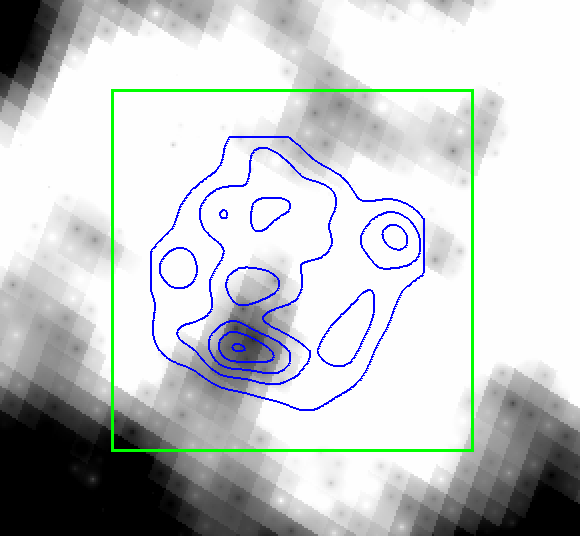}}
\caption{AKDE contours as in Figure~\ref{fig_over} overlaid on the \textit{Planck} 857 GHz map \citep{planckmaps}. The green box is $30\arcmin \times 30\arcmin$. The peak \textit{Planck} 857 GHz flux in the overdensity is 2.56 MJy/sr. Note: darker is brighter, as in previous figures. North is up, East is to the left.}
\label{fig_planck}
\end{figure}

We also identified the overdensity in the \textit{Planck}-HFI 857 GHz map, shown in Figure~\ref{fig_planck} as a low surface brightness feature (2.56 MJy/sr peak value measured from the map), hence its absence from the \textit{Planck} Early Release Compact Source Catalog (ERCSC, \citealt{planckcat}). The detection with \textit{Planck} is in support of the efficiency of the beam matching detection (\textit{Planck}-HFI 857 GHz beam FWHM is $5\arcmin$) although the overdensity is not well pronounced at 350 \micron\ (corresponding to the HFI 857 GHz band). This hints that the majority of the sources in the clump are brighter at 250 \micron\ and thus at redshift below 3, where \textit{Planck} is not that efficient for sources dominated by dust emission.

\section{Conclusions}
\label{sec:conclusions}

In this study we presented a serendipitous detection of a significant {($5.1\sigma$)} overdensity of 250 \micron\ detected star-forming galaxies. To quantify the significance we made use of an adaptive kernel density estimate of the galaxy distribution. {In fields with the same geometry and the same number but randomly distributed sources we found that only 2\% show similar peaks.} We do not see an overdensity of similar size or similar significance in four extragalactic control fields: GOODS-North, COSMOS, Lockman and UDS. 

The surface number density of the detected SPIRE 250 \micron\ galaxies in the overdensity region are significantly in excess with respect to the average galaxy surface density from the four control fields.  A similar surface density in a $4\arcmin$ region was found in the COSMOS field,  hence regions with similar excess over the background do exist, but they do not necessarily correspond to a significant density peak, e.g. in COSMOS. Our interpretation is that the sources in the COSMOS region are more homogeneously distributed within the $4\arcmin$ region, while they are more centrally concentrated in the overdensity that is discussed here.

Additional support for the presence of the overdensity comes from the 250 \micron\ number counts. There is a significant excess of sources with flux densities between 30 and 40 mJy in the overdensity compared to the control fields and the {published} SPIRE number counts. We note, however, that the excess in the 30-40 mJy bin may be due to the combined effects of source confusion and flux boosting, because of the closely packed sources in the clump.

Our photometric analysis, based on only the 3 SPIRE bands, makes use of a black-body redshift $z_\mathrm{BB}$. This is the best fit redshift of a modified black-body, with a fixed dust temperature and emissivity, to the three (at best) photometry measurements. The colour information and the SED shape are encapsulated in $z_\mathrm{BB}$. We have shown that the $z_\mathrm{BB}$ distribution exhibits a significant peak for the overdensity region. No similar peak in $z_\mathrm{BB}$ exists in the control fields. We showed that the galaxies within the peak have similar 250\micron/350\micron\ colours, as expected by the definition of $z_\mathrm{BB}$.

The reported overdensity can also be seen as a low surface brightness feature in the \textit{Planck} 857 GHz map, although too faint to be included in the {official public} point source catalogue. This emphasises the need for a parallel search of visual overdensities of dusty star-forming galaxies in fields observed with \textit{Herschel}. Such clumps of dusty star-forming galaxies are expected to exist (e.g. \citealt{negrello05}) but there are no reports up to now, mainly because these clumps are rare, and because there was a lack of large extragalactic fields observed in the far-infrared, prior to \textit{Planck} and \herschel.

All these results provide viable arguments that we may be seeing a real clump of dusty star-forming galaxies that may reside within the same large-scale structure. Moreover, the overdensity was identified in the field of a known protocluster: the Spiderweb. It lies close in projection (7 arcmin corresponding to a comoving distance of 10 Mpc at $z=2.0$) to the central protocluster massive and intensively star-forming radio galaxy MRC\,1138-26. A large-scale filament traced by H$\alpha$ emitters of a comoving size of $\gtrsim10$ Mpc and going from NW to SE was already reported in the Spiderweb field (see \citealt{koyama13}), although the narrow band filter observations did not cover as far as the overdensity region. 

Nevertheless, the proximity to the protocluster raises the valid and scientifically attractive question: is the clump part of the same large-scale complex? Due to the scarce multi-wavelength data, poor statistics of SPIRE-detected spectroscopically confirmed Spiderweb members, and the large uncertainties in the colour-colour, colour-magnitude and the crude FIR redshift estimate, we cannot provide any {conclusive} evidence to support or discard this hypothesis. Moreover, there are known protoclusters in the control fields and none of them correspond to any visually compelling overdensities of 250 \micron\ sources.  Consequently this hypothesis remains highly speculative and extensive ground-based photometric and spectroscopic follow-up observations, which we are currently pursuing, will be needed in order to confirm or discard it.

\section*{Acknowledgments}
{We would like to thank the anonymous referee for the comprehensive comments that helped to greatly improve the paper.} \\\herschel\ is an ESA space observatory with science instruments provided by European-led Principal Investigator consortia and with important participation from NASA. SPIRE has been developed by a consortium of institutes led by Cardiff University (UK) and including Univ. Lethbridge (Canada); NAOC (China); CEA, LAM (France); IFSI, Univ. Padua (Italy); IAC (Spain); Stockholm Observatory (Sweden); Imperial College London, RAL, UCL-MSSL, UKATC, Univ. Sussex (UK); and Caltech, JPL, NHSC, Univ. Colorado (USA). This development has been supported by national funding agencies: CSA (Canada); NAOC (China); CEA, CNES, CNRS (France); ASI (Italy); MCINN (Spain); SNSB (Sweden); STFC (UK); and NASA (USA).\\ This publication makes use of data products from the Wide-field Infrared Survey Explorer, which is a joint project of the University of California, Los Angeles, and the Jet Propulsion Laboratory/California Institute of Technology, funded by the National Aeronautics and Space Administration.

\label{lastpage}

\end{document}